\documentclass[journal]{IEEEtran}

\usepackage{amsmath}
\usepackage{amssymb}
\usepackage{amsthm}
\usepackage{amsfonts}
\usepackage{graphicx}
\usepackage{subfigure}
\usepackage{nicefrac}
\usepackage{lipsum}
\usepackage{nccmath}
\usepackage{psfrag}
\usepackage{cite}
\usepackage{color}
\usepackage{bm}

\usepackage{tikz}
\usetikzlibrary{arrows,positioning}
\usetikzlibrary{calc}
\usetikzlibrary{decorations}
\usetikzlibrary{decorations.pathreplacing}
\usetikzlibrary{shapes.geometric}

\DeclareMathOperator*{\argmin}{arg\,min}
\DeclareMathOperator*{\argmax}{arg\,max}

\usepackage{lipsum}

\newcommand\blfootnote[1]{%
  \begingroup
  \renewcommand\thefootnote{}\footnote{#1}%
  \addtocounter{footnote}{-1}%
  \endgroup
}

\newcommand{\I}{\boldsymbol{\mathcal{I}}}
\newcommand{\rank}{\mbox{rank}}
\newcommand{\Ns}{N_{\!s}}
\newcommand{\Nw}{N_{\!\omega}}
\newcommand{\ys}{\mathbf{y}_{\!s}}
\newcommand{\yw}{\mathbf{y}_{\!\omega}}

\begin{document}
%\title{Inertial Sensor Arrays -- Information Fusion, Dynamic Range Extension, and Cram\'{e}r-Rao Bound}
%\title{Inertial Sensor Arrays -- Maximum Likelihood Based Measurement Fusion and Cram\'{e}r-Rao Bound}
\title{Inertial Sensor Arrays, Maximum Likelihood, and Cram\'{e}r-Rao Bound}
%\title{MIMU, Maximum Likelihood, and Cram\'{e}r-Rao Bound}
%\title{Maximum Likelihood Based Information Fusion for Inertial Sensor Arrays}

\author{Isaac~Skog~\IEEEmembership{Member,~IEEE}, John-Olof Nilsson~\IEEEmembership{Member,~IEEE}, Peter~H\"{a}ndel~\IEEEmembership{Senior Member,~IEEE}, and Arye Nehorai~\IEEEmembership{Fellow,~IEEE}% <-this % stops a space
%\thanks{Copyright (c) 2016 IEEE. Personal use of this material is permitted. However, permission to use this material for any other purposes must be obtained from the IEEE by sending a request to pubs-permissions@ieee.org.}
\thanks{I. Skog, J-O. Nilsson, and P. H\"{a}ndel are with the Department of Signal Processing, ACCESS Linnaeus Centre, KTH Royal Institute of Technology, Stockholm, Sweden. (e-mail:
skog@kth.se, jnil02@kth.se, ph@kth.se).}
\thanks{A. Nehorai is with the Preston M. Green Department of Electrical and Systems
Engineering, Washington University in St. Louis, St. Louis, MO 63130
USA (e-mail: nehorai@ese.wustl.edu).}
\thanks{The work of I. Skog, J. Nilsson, and P. H\"{a}ndel has partly been supported by the Swedish Governmental Agency for Innovation Systems (VINNOVA).}
\thanks{The work of A. Nehorai was support by AFOSR Grant No. FA9550-11-1-0210.}}% <-this % stops a space

\maketitle

\begin{abstract}
A maximum likelihood estimator for fusing the measurements in an inertial sensor array is presented. The maximum likelihood estimator is concentrated and an iterative solution method is presented for the resulting low-dimensional optimization problem. The Cram\'{e}r-Rao bound for the corresponding measurement fusion problem is derived and used to assess the performance of the proposed method, as well as to analyze how the geometry of the array and sensor errors affect the accuracy of the measurement fusion. The angular velocity information gained from the accelerometers in the array is shown to be proportional to the square of the array dimension and to the square of the angular speed.
% In contrast the information about the angular acceleration is shown to reduce with increasing angular speed. %Simulations indicates that the proposed fusion method is statistically efficient, i.e., it attains the Cram\'{e}r-Rao bound even for a finite number of sensors.
In our simulations the proposed fusion method attains the Cram\'{e}r-Rao bound and outperforms the current state-of-the-art method for measurement fusion in accelerometer arrays. Further, in contrast to the state-of-the-art method that requires a 3D array to work, the proposed method also works for 2D arrays. The theoretical findings are compared to results from real-world experiments with an in-house developed array that consists of 192 sensing elements.
%\lipsum[1]
\end{abstract}

%\begin{IEEEkeywords}
%IEEEtran, journal, \LaTeX, paper, template.
%\end{IEEEkeywords}

% \fi
\IEEEpeerreviewmaketitle

\ifCLASSOPTIONcaptionsoff
  \newpage
\fi

\section{Introduction}
\begin{figure}[t]
\centering
\begin{tikzpicture}
   \node[anchor=south west,inner sep=0] (image) at (0,0) {\includegraphics[width=0.6\columnwidth]{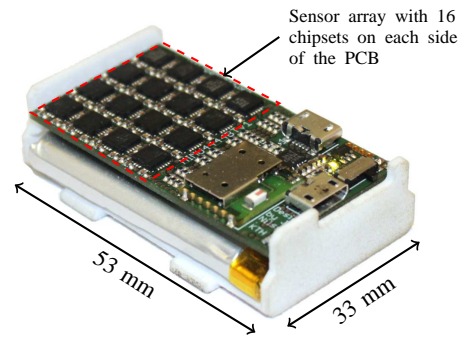}};
    \begin{scope}[x={(image.south east)},y={(image.north west)}]
        \draw[<->,black,thick] (-0.01,0.46) -- (0.59,-0.03) node[pos=0.5,below,rotate=-32]{\small 53 mm};
        \draw[<->,black,thick] (0.67,-0.01) -- (1,0.27) node[pos=0.5,below,rotate=34]{\small 33 mm};
        \draw[red,dashed,thick] (0.36,0.93) -- (0.65,0.73) -- (0.32,0.47)--(0.03,0.70)--(0.36,0.93); %node[pos=1/2,below,rotate=34]{\small kkk};
        \draw[<-,black,thick] (0.51,0.82) -- (0.65,0.95) node[right,black,text width=25mm,align=left]{\baselineskip=5pt \scriptsize{Sensor array with 16 chipsets on each side of the PCB}\par};
    \end{scope}
%\node at (0mm,0mm) [anchor=north west,inner sep=0]{\includegraphics[width=0.7\columnwidth]{MIMU4444BT_module_open.eps}};
%\draw[|-|,semithick] (0,-49.3mm-3mm) -- (5.6mm,-49.3mm-3mm) node[midway,below] {\footnotesize 26.6 [mm]};
%\draw[|-|,semithick] (0-3mm,0) -- (0-3mm,-49.3mm) node[midway,left,anchor=south,rotate=90] {\footnotesize 49.3 [mm]};
\end{tikzpicture}
  \caption{An in-house designed embedded system with an inertial sensor array, which is available under an open-source licence. The array consists of 32 MPU9150 Invensense inertial sensor chipsets, each containing an accelerometer triad, a gyroscope triad, and a magnetometer triad.}\label{F:array}
  %\caption{Top and bottom side of an in-house designed inertial sensor array, available under an open-source licence. The array consist of 32 individual MPU9150 Invensense inertial sensor chipsets, each containing an accelerometer triad, a gyroscope triad, and a magnetometer triad.}\label{F:array} %The communication of with all the arrays is handeled in parallel by the AVR32UC3C0512 microcontroller from Atmel.}\label{F:array}
\end{figure}

\PARstart{M}{otion} sensing is an essential capability in many systems to achieve a high level of autonomy. Nowadays, inertial motion sensors are ubiquitous in everything from industrial manufacturing equipment to consumer electronic devices. This widespread usage of inertial sensors has become possible thanks to the last decade's micro-electrical-mechanical-system technology development, which has revolutionized the inertial sensor industry~\cite{Shaeffer2013}. Today, inertial sensors can be manufactured at unprecedented volumes and at low prices~\cite{Perlmutter2012}; six degrees-of-freedom inertial sensor assemblies where sold to large-volume customers at less than a dollar in 2013 \cite{YoleReport}. Even though development has pushed the performance boundaries of the inertial sensor technology, contemporary ultralow-cost sensors still cannot fully meet the needs of many applications; especially applications where the sensor signals have to be integrated over time. These applications still suffer from the bias instability, nonlinearities, and thermal instability of the contemporary ultralow-cost sensors~\cite{Shaeffer2013}. However, by capitalizing on the decreasing price, size, and power-consumption of the sensors, it is now feasible to construct arrays with hundreds of sensing elements, and digitally process these measurements to create virtual high-performance sensors. The benefit of using an array of sensors is not only an increased measurement accuracy, but also an increased operation reliability thanks to the possibility of carrying out sensor fault detection and isolation~\cite{Bittner2014,Song2015}. Further, from the redundant measurements the covariance of the measurement errors can be estimated and used to determine the reliability of the measurements \cite{Waegli2010}. Moreover, as will be shown, the angular acceleration can be directly estimated and for some array configurations the dynamic measurement range can be extended beyond that of the individual sensors by utilizing the spatial separation of the sensors. An example of an in-house developed embedded system holding an inertial sensor array constructed from contemporary ultralow-cost sensors is shown in Fig.~\ref{F:array}. The system has 192 inertial sensing elements, a micro-controller for information fusion, and a Bluetooth interface for communication. Refer to {\emph{www.openshoe.org}} and \cite{SNH2014a} for details.

%\begin{figure}[t]
%\centering
%\includegraphics[width=0.6\columnwidth]{MIMU4444BT_module_open.eps}
%  \caption{In-house designed embedded system with an inertial sensor array, available under an open-source licence. The array consist of 32 individual MPU9150 Invensense inertial sensor chipsets, each containing an accelerometer triad, a gyroscope triad, and a magnetometer triad.}\label{F:array}
%\end{figure}

\blfootnote{
  \normalsize{\emph{Reproducible research: The layout files and software for the inertial sensor array used in the experiments are available under an open-source licence at www.openshoe.org.}}}

Inertial sensors are primarily used to measure the motion of objects. Since a rigid body in a three-dimensional space has six degrees-of-freedom, three rotational and three translational, the motion of an object is commonly measured using a sensor assembly of three gyroscopes and three accelerometers, a.k.a. an inertial measurement unit (IMU). However, as a rigid object's motion can also be described in terms of the translational motion of three non-collinear points on the object, it is also possible to measure the motion using only an assembly of spatially separated accelerometers, a.k.a. a gyroscope-free IMU; the object's rotational motion is then described by the relative displacement of the points. Thus, with an inertial sensor assembly that consists of both gyroscopes and spatially distributed accelerometers, a.k.a. an inertial sensor array, rotational information can be obtained both from the gyroscopes and the accelerometers. The question is, how should the measurements in such sensor array be fused?

Array signal processing and measurement fusion in the context of sensor arrays measuring wave fields is a long-studied subject in the signal processing literature~\cite{Krim1996}, and the within the area developed methods have been applied to various types of sensor arrays such as antenna arrays~\cite{Stoica1989,Viberg1991},  magnetic sensor arrays~\cite{Nehorai1994}, acoustic sensor arrays~\cite{Nehorai1994b,Abdi2009,Zou2009}, and chemical sensor arrays \cite{Jeremic2000}. However, since an inertial sensor array do not measure a wave field, but something that can be referred to as a force field, the signal models and methods in the array processing literature are not directly applicable to the inertial sensor array measurement fusion problem. Nevertheless, based on results from classical mechanics about forces in rotating coordinate frames a signal model for the inertial sensor array measurements can be formulated, and results from the field of estimation theory applied to derive an measurement fusion method. Accordingly, we will propose a maximum likelihood approach for fusing the measurements in an inertial sensor array consisting of multiple accelerometer and gyroscope triads. With suitable projection vectors, the approach can also be generalized to a system of single sensitivity axis sensor units.

% but something that can be viewed as a force field created by the difference in motion trajectory of a set of non-collocated points on a solid body, \cite{Krim1996,Stoica1989,Viberg1991,Nehorai1994,Nehorai1994b,Abdi2009,Zou2009,Jeremic2000}
%In this paper, we will propose a maximum likelihood approach for fusing the measurement information in a sensor assembly consisting of multiple accelerometer and gyroscope triads, a.k.a an inertial sensor array. With suitable projection vectors, the approach can also be generalized to a system of single sensitivity axis sensor units.

\subsection{State-of-the-art techniques}
Next, a brief summary of existing work on inertial sensors arrays is presented. For an in-depth review of the topic the reader is referred to the literature survey presented in~\cite{Nilsson2016}.

In the 1960s, it was shown that the specific force and angular acceleration of an object can be estimated with an assembly of nine accelerometers\footnote{With six accelerometers the angular velocity cannot be estimated from the measurements at a single time instant, however, a differential equation describing the rotation dynamics can be posed by which the angular velocity can be tracked over time, see e.g. \cite{Williams2013,Tan2005}.}\cite{Schuler1967}; the angular velocity can only be estimated up to a sign ambiguity since the centrifugal force depends quadratically on the angular speed~\cite{Schopp2014,Sukkarieh2000,Chatterjee2015}. Though the motion of an object can be tracked using only the estimated specific force and angular acceleration, the increased sensitivity to measurement errors caused by the extra integration needed to calculate the orientation of the tracked object renders it as an intractable solution. Therefore, a variety of methods to resolve the sign ambiguity of the angular velocity estimates by tracking the time development of the estimated quantities have been proposed \cite{Naseri2014,He2012,Madgwick2013,Park2011,Schopp2010,Parsa2004,Cardou2008,Schopp2014}.

Instead of directly estimating the angular velocity and angular acceleration from the accelerometer measurements, it is common to first estimate the angular acceleration tensor (or variations thereof), and then calculate the angular velocity and angular acceleration \cite{Qin2009,Parsa2004}. To simplify the estimation of the angular acceleration tensor, the fact that it only has six-degrees of freedom is generally neglected, so that it can be estimated via a linear transformation of the accelerometer measurements, see e.g. \cite{Parsa2007,Qin2009,He2012,Madgwick2013,Park2011,Schopp2010,Williams2013,Liu2014,Edwan2011} or the Appendix. The advantages of this approach are its simplicity and that it is straight-forward, using the least squares method (see e.g. \cite{Parsa2004}), to include the measurements from additional redundant accelerometers.  The disadvantages are mainly caused by the neglected constraints on the tensor, these are as follows: (i) a minimum of twelve, instead of nine, accelerometers are needed for the estimation problem to be well defined \cite{Qin2009}; (ii) the estimation accuracy is reduced; and (iii) the locations of the sensors must span a 3D space instead of a 2D space~\cite{Madgwick2013}. The requirement that the locations of the sensors must span a 3D space significantly increases the size of such a system and causes a problem if a sensor array is to be constructed on a printed circuit board.

Since the angular velocity of all points on a rigid object are the same, no additional information, except that from redundant measurements, is obtained from having several spatially distributed gyroscopes in an inertial sensor array. Still, a non-negligible reduction in the measurement errors can be obtained from the redundant measurements, see e.g. the Allan variance analysis in \cite{SNH2014a}. A few methods to fuse the measurements from multiple gyroscopes using different filter frameworks and signal models are described in \cite{Jiang2012,Yuksel2011,Xue2012}.

With respect to inertial arrays that consist of multiple IMUs, the measurement fusion problem has mainly been studied in the framework of global navigation satellite system aided inertial navigation systems \cite{Waegli2010,Bancroft2011,Jafari2014,Guerrier2009,Bancroft2009}. In the literature, the proposed information fusion approaches can be broadly grouped into two categories. In the first category, see e.g. \cite{Waegli2010,Bancroft2011,Jafari2014,Guerrier2009}, the IMU measurements are fused before they are used in the inertial navigation systems; commonly a weighted average is used and the spatial separation of the sensors is neglected. In the second category, see e.g. \cite{Bancroft2009}, they are fused after being processed by several parallel inertial navigation systems. Refer to \cite{Bancroft2011} and the references therein for a discussion on the pros and cons of the two approaches, as well as an evaluation of different measurement fusion methods, including a few where the spatial separation of the sensors are taken into account.

\subsection{Presented work and findings}
In this paper, the problem of fusing the measurements from an array of accelerometer and gyroscope triads is posed as a parameter estimation problem for the first time. Results from the field of estimation theory are used to derive a maximum likelihood based measurement fusion method. The Cram\'{e}r-Rao bound (CRB) for the corresponding estimation problem is also derived and used to evaluate the performance of the proposed measurement fusion method. Simulations show that the proposed method attains the CRB and outperforms the current state-of-the-art method for measurement fusion in accelerometer arrays. Further, by studying the properties of the signal model it is shown that necessary and sufficient conditions for the measurement fusion problem to be well conditioned is that the array consists of at least one gyroscope triad and three accelerometer triads whose locations are non-collinear; current state-of-the-art information fusion methods require 3D arrays. Moreover, an analysis of the CRB reveals that: (i) the angular velocity information gained from the accelerometers is proportional to the square of the array dimensions and to the square of the angular speed; (ii) the accuracy with which the angular acceleration can be estimated decreases with the angular velocity; and (iii) there exists no measurement fusion method that can calculate an unbiased, finite variance, estimate of the angular velocity for small angular velocities from an array of only accelerometer. To support the theoretical findings, the accuracy of the proposed measurement method is also experimentally evaluated. The experimental results show that proposed method works, but it also shows that there is a discrepancy between the theoretical and experimental results. The discrepancy between the theoretical and experimental results are likely primarily due to uncertainties in the location of the sensing elements of the real-world array. To summarize, the proposed information fusion method outperforms the current state-of-the-art method and also works for 2D arrays, allowing for smaller and cheaper senor arrays to be constructed. However, more research on sensor array calibration is needed before the inertial sensor arrays can reach their full potential in practice.

\section{Measurement fusion}\label{S:Information fusion}
%\textcolor{red}{Array signal processing in the context of source localization based upon a wave field across an array of sensors is a long-studied subject \cite{Krim1996}. The within the area developed parameter estimation and information fusion methods have been applied to various types of sensor arrays such as antenna arrays~\cite{Stoica1989,Viberg1991},  magnetic sensor arrays~\cite{Nehorai1994}, acoustic sensor arrays~\cite{Nehorai1994b,Abdi2009,Zou2009}, and chemical sensor arrays \cite{Jeremic2000}. However, since an inertial sensor array don't measure a wave field, but something that can be viewed as a force field created by the difference in motion trajectory of a set of non-collocated points on a solid body, traditional array processing methods cannot be applied. The objective is also not to estimate the location of a signal source, but the motion of the non-stationary array. Therefore, an estimator to fuse the measurement from an inertial sensor array will be presented. The derivation of the estimator takes its starting point in the, from classical mechanics derived, relationship between forces in rotating coordinate frames.}

%\textcolor{red}{Consider an inertial sensor array consisting of $\Ns$ accelerometer triads and $\Nw$ gyroscope triads. The maximum likelihood estimator that estimates the specific force, angular velocity, and angular acceleration of the array from the sensors measurements can then be derived as follows.}

In this section a maximum likelihood estimator to fuse the measurement from an inertial sensor array consisting of $\Ns$ accelerometer triads and $\Nw$ gyroscope triads will be presented. The derivation of the estimator takes its starting point in the, from classical mechanics obtained, relationship between forces in rotating coordinate frames.

\subsection{Accelerometer signal model}
\begin{figure}[t!]
\centering
\begin{tikzpicture}
\node at (16.1mm,0mm) [opacity=1] {\includegraphics[width=70mm]{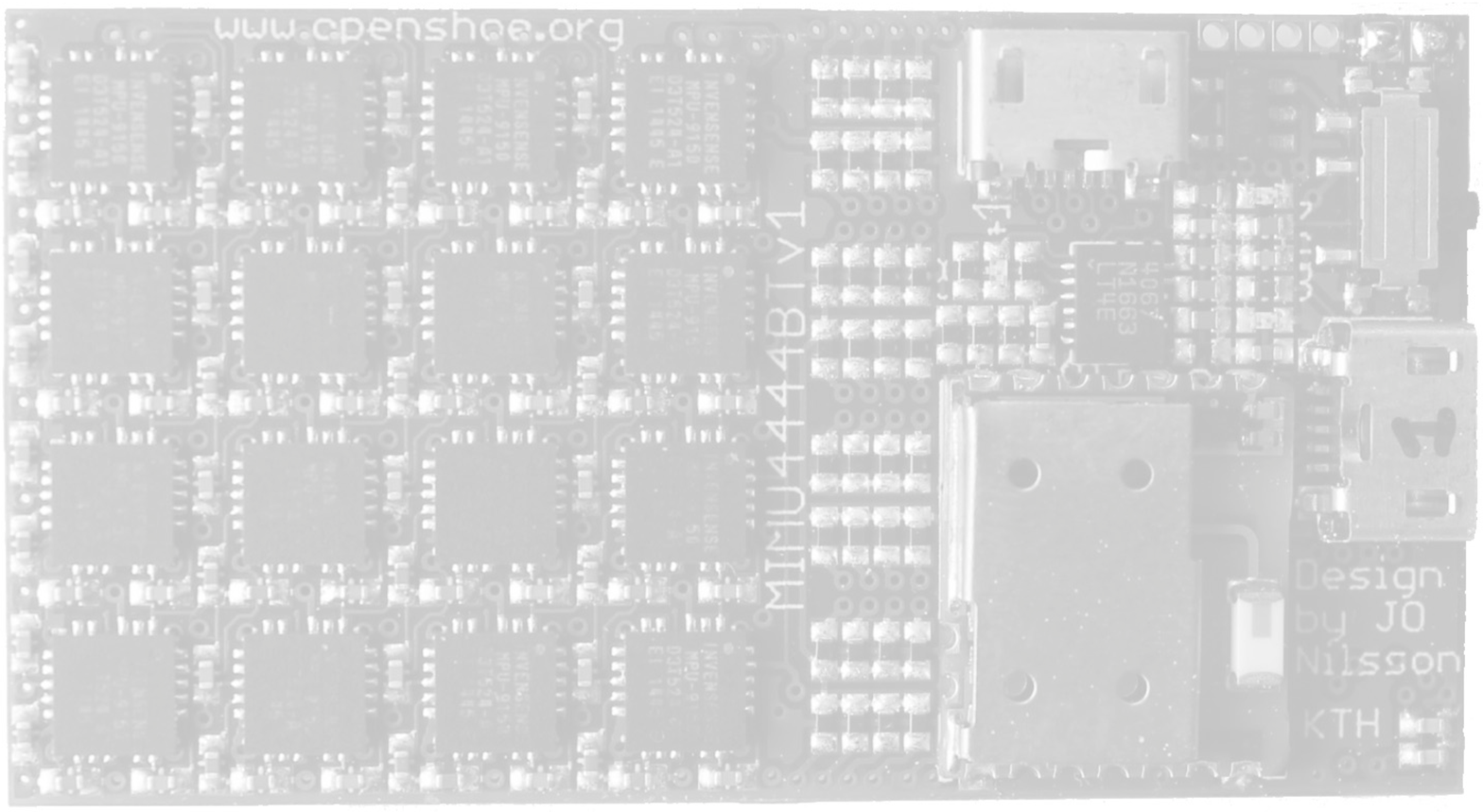}};
\foreach \x in {0,1,2,3} {
  \foreach \y in {0,1,2,3} {
    \node at ($(\x*9mm,\y*9mm)+(-13.55mm,-13.35mm)$) [rectangle,draw=white,minimum size=5.5mm,inner sep=0mm,fill=none] {\tiny\color{white} IMU};
  }
}
\draw[->] (0,0) -- (7mm,0) node[pos=1.3] {\small $x$};
\draw[->] (0,0) -- (0,7mm) node[pos=1.3] {\small $y$};
\draw[->,red] (0,0) -- (-10mm,3mm)  node[pos=1.2] {\color{blue}\small $\boldsymbol{s}$};
\node[circle,fill=black,minimum size=2mm,inner sep=0mm] {};
\node at (0mm,-3mm) {\small Origin of array coordinate frame};
\node at (-15mm,6mm) {\small Specific force};
\draw[|->,densely dashed,black!88] (70:4mm) arc (70:185:4mm);
\draw[densely dashed,black!88] (19.1mm,0) arc (0:90:19.1mm);
\node at (135:6mm) {\color{blue}\small $\boldsymbol{\omega}$};
\draw (0,0) -- (13.5mm,13.5mm) node[pos=0.45,above=1mm] {\color{blue}\small $\boldsymbol{r}_i$};
\draw[->] (13.5mm,13.5mm) -- +(6mm,0) node[pos=1.3] {\small $x$};
\draw[->] (13.5mm,13.5mm) -- +(0,6mm) node[pos=1.3] {\small $y$};
\draw[->,red] (13.5mm,13.5mm) -- +(135:6mm);
\node at (4mm,17.5mm) {\color{blue}\small $\boldsymbol{\omega}\times\boldsymbol{r}_i$};
\node at (3.5mm,21mm) {\small Euler force};
\draw[->,red] (13.5mm,13.5mm) -- +(45:6mm);
\node at (28mm,20mm) {\small Centrifugal force};
\node at (28mm,17mm) {\color{blue}\small $\boldsymbol{\omega}\times(\boldsymbol{\omega}\times\boldsymbol{r}_i)$};
\node at (13.5mm,13.5mm) [rectangle,fill=black,minimum size=1.75mm,inner sep=0mm] {};
\node at (27mm,10mm) {\small $i$:th accelerometer triad};
\end{tikzpicture}
  \caption{An illustration of the decomposition of the force sensed by an accelerometers in an inertial sensor array, overlaid on a picture of the printed circuit board of the inertial sensor array shown in Fig.~\ref{F:array}. The specific force sensed by the $i$:th accelerometer triad is the sum of the specific force at the origin of the array coordinate frame, the centrifugal force, and the Euler force.
  }\label{F:forces}
\end{figure}

Basic kinematics dictates that the specific force in one point of a rotating coordinate frame may be decomposed into that of another point, a centrifugal term, and an Euler term. Specifically, as illustrated in Fig.~\ref{F:forces}, the specific force $\mathbf{s}_i$, sensed by the $i$:th accelerometer triad located at position $\mathbf{r}_i$ in the array coordinate frame and which sensitivity axis are aligned with the coordinate axis of the array frame\footnote{A calibration method to align the coordinate axis of the sensors in an inertial sensor array can be found in \cite{NSH2014}.}, is given by \cite[p.~90]{Jekeli2001}
\begin{equation}\label{E:force eq}
\mathbf{s}_i=\mathbf{s}+\underbrace{\boldsymbol{\omega}\times(\boldsymbol{\omega}\times\mathbf{r}_i)}_{\mbox{\footnotesize Centrifugal force}}
+\underbrace{\dot{\boldsymbol{\omega}}\times\mathbf{r}_i}_{\mbox{\footnotesize{Euler force}}}.\end{equation}
Here $\mathbf{s}$ denotes the specific force sensed at the origin of the array frame. Further, $\boldsymbol{\omega}$ and $\dot{\boldsymbol{\omega}}$ denote the array frame's angular velocity and angular acceleration w.r.t. inertial space, respectively. Moreover, $\mathbf{a}\times\mathbf{b}$ denotes the cross product between vector $\mathbf{a}$ and $\mathbf{b}$. This relates the measurements of different accelerometers of known locations within the array to a common specific force $\mathbf{s}$ via the angular velocity $\boldsymbol{\omega}$ and acceleration $\dot{\boldsymbol{\omega}}$ of the array.

%Fig.~\ref{F:forces} shows an illustration of the relationship between the forces in (\ref{E:force eq}).

%\begin{figure}[t!]
%  \centering
%  \psfrag{w}[c][c][0.8][0]{$\boldsymbol{\omega}$}
%  \psfrag{e1}[c][c][0.8][0]{Euler force}
%  \psfrag{e2}[c][c][0.8][0]{$\dot{\boldsymbol{\omega}}\times\mathbf{r}_i$}
%  \psfrag{c1}[c][c][0.8][0]{Centrifugal force}
%  \psfrag{c2}[c][c][0.8][0]{$\boldsymbol{\omega}\times(\boldsymbol{\omega}\times\mathbf{r}_i)$}
%  \psfrag{r}[c][c][0.8][0]{$\mathbf{r}_i$}
%  \psfrag{x}[c][c][0.8][0]{$x$}
%  \psfrag{y}[c][c][0.8][0]{$y$}
%  \psfrag{o}[c][c][0.8][0]{Origin of array coordinate frame}
%  \psfrag{s}[l][l][0.8][0]{$i$:th accelerometer triad}
%  \psfrag{sp1}[c][c][0.8][0]{Specific force}
%    \psfrag{sp2}[c][c][0.8][0]{$\mathbf{s}$}
%  \includegraphics[width=0.5\columnwidth]{forces3.eps}
%  \caption{An illustration of the forces sensed by the accelerometers in the array. The specific force sensed by the $i$:th accelerometer triad is the sum of the specific force at the origin of the array coordinate frame, the centrifugal force, and the Euler force.}\label{F:forces}
%\end{figure}

Next, let $\boldsymbol{\Omega}_\mathbf{a}$ denote the skew-symmetric matrix representations of the three element vector $\mathbf{a}$ for which $\boldsymbol{\Omega}_\mathbf{a}\mathbf{b}=\mathbf{a}\times\mathbf{b}$. Then (\ref{E:force eq}) may be rewritten in terms of matrix multiplications as
\begin{equation}\label{E:force eq 2}
\begin{split}
\mathbf{s}_i&=\mathbf{s}+\boldsymbol{\Omega}_{\boldsymbol{\omega}}\boldsymbol{\Omega}_{\boldsymbol{\omega}}\mathbf{r}_i+\dot{\boldsymbol{\Omega}}_{\boldsymbol{\omega}}\mathbf{r}_i\\
&=\mathbf{s}+\boldsymbol{\Omega}_{\boldsymbol{\omega}}^2\mathbf{r}_i-\boldsymbol{\Omega}_{\mathbf{r}_i}\dot{\boldsymbol{\omega}},
\end{split}
\end{equation}
where in the second equality it was used that $\mathbf{a}\times\mathbf{b}=-\mathbf{b}\times\mathbf{a}$.
The measurement vector $\ys$, consisting of the measurements from the accelerometers in the array, can thus be modeled as
\begin{equation}\label{E:sig model accelerometer}
\ys=\mathbf{h}_s(\boldsymbol{\omega})+\mathbf{H}_s\boldsymbol{\phi}+\mathbf{n}_s,
\end{equation}
where %$\mathbf{H}_s=\begin{bmatrix}\mathbf{G} & \mathbf{1}_{\Ns}\!\otimes\mathbf{I}_3\\ \end{bmatrix}$ and

\begin{equation*}
\mathbf{h}_s(\boldsymbol{\omega})\!=\!\begin{bmatrix}\boldsymbol{\Omega}_{\boldsymbol{\omega}}^2\mathbf{r}_1\\
\vdots\\
\boldsymbol{\Omega}_{\boldsymbol{\omega}}^2\mathbf{r}_{\Ns}\\\end{bmatrix}\!,\,\,
\mathbf{H}_s=\begin{bmatrix}\mathbf{G} & \mathbf{1}_{\Ns}\!\otimes\mathbf{I}_3\\ \end{bmatrix}\!,\,\,
\mathbf{G}\!=\!\begin{bmatrix}
-\boldsymbol{\Omega}_{\mathbf{r}_1}\\
\vdots\\
-\boldsymbol{\Omega}_{\mathbf{r}_{\Ns}}
\end{bmatrix}\!,
\end{equation*}
and $\boldsymbol{\phi}=\begin{bmatrix} \dot{\boldsymbol{\omega}}^\top & \mathbf{s}^\top \end{bmatrix}^\top$. Here $\mathbf{1}_n$ and $\mathbf{I}_n$ denote a column vector of length $n$ with all of the entries equal to one and an identity matrix of size $n$, respectively. Further, $\mathbf{n}_s$ denotes the measurement error of the accelerometers.

\subsection{Gyroscope signal model}
The angular velocity sensed by the gyroscopes is independent of their location in the array. Consequently, the measurement vector $\yw$, consisting of the measurements from all of the gyroscopes in the array, can be modeled as
\begin{equation}\label{E:sig model gyroscopes}
\yw=\mathbf{h}_\omega(\boldsymbol{\omega})+\mathbf{n}_\omega,
\end{equation}
where $\mathbf{h}_\omega(\boldsymbol{\omega})=\mathbf{1}_{\Nw}\!\otimes\boldsymbol{\omega}$, and $\mathbf{n}_\omega$ denotes the measurement error of the gyroscopes. That is, all gyroscopes measures the same angular velocity and are disturbed by additive measurement error.

\subsection{Array signal model}
Concatenating all of the measurements into a single vector, the signal model for the full array becomes
\begin{equation}\label{E:sig model}
\mathbf{y}=\mathbf{h}(\boldsymbol{\omega})+\mathbf{H}\,\boldsymbol{\phi}+\mathbf{n},
\end{equation}
where
%\rightarrow
\begin{equation*}
\mathbf{y}\!=\!
\begin{bmatrix}
\ys\\
\yw
\end{bmatrix}
\,\,
\mathbf{h}(\boldsymbol{\omega})=
\begin{bmatrix}
\mathbf{h}_s(\boldsymbol{\omega})\\
\mathbf{h}_\omega(\boldsymbol{\omega})
\end{bmatrix}
\,\,
\mathbf{H}\!=\!
\begin{bmatrix}
\mathbf{H}_s\\
\mathbf{0}_{3\Nw,6} %\mathbf{0}
\end{bmatrix}
\,\,
\mathbf{n}\!=\!
\begin{bmatrix}
\mathbf{n}_s\\
\mathbf{n}_\omega\\
\end{bmatrix}.
\end{equation*}
Here $\mathbf{0}_{n,m}$ denotes a zero matrix of size $n$ by $m$. Thus, the signal model (\ref{E:sig model}) for the inertial sensor array consists of a non-linear part $\mathbf{h}(\boldsymbol{\omega})$, depending only on the angular velocity $\boldsymbol{\omega}$, and a linear part $\mathbf{H}\,\boldsymbol{\phi}$,  depending only on the specific force $\mathbf{s}$ and angular acceleration $\dot{\boldsymbol{\omega}}$. This separation of dependencies will turn out to be useful in the estimation of the specific force and the angular velocity.

\subsection{Identifiability conditions}
When $\Nw=0$, i.e., the array consists of only accelerometers, then $\mathbf{h}(\boldsymbol{\omega})\equiv\mathbf{h}_s(\boldsymbol{\omega})=\mathbf{h}_s(-\boldsymbol{\omega})$ and the parameters in the signal model is not identifiable.
%Noteworthy is, that when $\Nw=0$, i.e., the array consists of only accelerometers, then $\mathbf{h}(\boldsymbol{\omega})\equiv\mathbf{h}_s(\boldsymbol{\omega})=\mathbf{h}_s(-\boldsymbol{\omega})$ and the parameters in the signal model is not identifiable.
Further, $\mathbf{H}$ will only have full rank if the array has at least three accelerometer triads whose locations span a 2D space. To see this, assume without loss of generality that the origin of the array coordinate system is defined so that $\mathbf{r}_1=\mathbf{0}_{3,1}$, then

\begin{equation}
\begin{split}
  \rank(\mathbf{H})&=\rank(\!\begin{bmatrix}  \mathbf{0}_{3,3} & \mathbf{I}_3\\
  -\boldsymbol{\Omega}_{\mathbf{r}_2} \!&\! \mathbf{I}_3\\
  \vdots & \vdots\\
  -\boldsymbol{\Omega}_{\mathbf{r}_{\Ns}} \!&\! \mathbf{I}_3\\
  \mathbf{0}_{3\Nw,3} \!&\! \mathbf{0}_{3\Nw,3}\\
  \end{bmatrix}\!)=3+\rank(\!\begin{bmatrix}
  -\boldsymbol{\Omega}_{\mathbf{r}_2}\\
  \vdots\\
  -\boldsymbol{\Omega}_{\mathbf{r}_{\Ns}}\\
  \end{bmatrix}\!)\nonumber.
  \end{split}
\end{equation}
Next, note that for $\mathbf{r}_i\neq\mathbf{0}$, $\rank(\boldsymbol{\Omega}_{\mathbf{r}_i})=2$ and the null space vector of $\boldsymbol{\Omega}_{\mathbf{r}_i}$ is $\mathbf{r}_i$. Thus, only if there are two accelerometer triads whose locations $\mathbf{r}_i$ ($\mathbf{r}_i\neq\mathbf{0}$) and $\mathbf{r}_j$ ($\mathbf{r}_j\neq\mathbf{0}$) are such that $\mathbf{r}_i\neq\beta\,\mathbf{r}_j$ $\forall\beta\in\mathbb{R}$, then $\rank(\mathbf{H})=6$. That is, $\mathbf{H}$ has full rank if $\Ns\geq3$ and $\rank(\begin{bmatrix} \mathbf{r}_1&\!\hdots\!&\mathbf{r}_{\Ns} \end{bmatrix})\geq 2$. Since the angular velocity is directly measured by the gyroscopes this implies that necessary and sufficient conditions for the parameters in the signal model to be identifiable is that: (i) the array has at least one gyroscope triad, and (ii) at least three accelerometer triads whose locations span a 2D space.

\subsection{Maximum likelihood  measurement fusion}
Assuming the measurement error $\mathbf{n}$ to be zero-mean and Gaussian distributed with the known covariance matrix $\mathbf{Q}$, the log-likelihood function for the signal model (\ref{E:sig model}) is given by

\begin{equation}\label{E:LLH}
  L(\boldsymbol{\omega},\boldsymbol{\phi})=-\frac{1}{2}\|\mathbf{y}-\mathbf{h}(\boldsymbol{\omega})-\mathbf{H}\boldsymbol{\phi}\|^2_{\mathbf{Q}^{-1}}+c,
\end{equation}
where $c$ is a constant and $\|\mathbf{a}\|^2_\mathbf{M}=\mathbf{a}^\top\mathbf{M}\mathbf{a}$. The maximum likelihood estimator for $\{\boldsymbol{\omega},\boldsymbol{\phi}\}$ is thus given by
\begin{equation}\label{E:MLE}
\begin{split}
\{\hat{\boldsymbol{\omega}},\hat{\boldsymbol{\phi}}\}&=\argmax_{\boldsymbol{\omega},\boldsymbol{\phi}}\left(L(\boldsymbol{\omega},\boldsymbol{\phi})\right)\\&=\argmin_{\boldsymbol{\omega},\boldsymbol{\phi}}\left(\|\mathbf{y}-\mathbf{h}(\boldsymbol{\omega})-\mathbf{H}\boldsymbol{\phi}\|^2_{\mathbf{Q}^{-1}}\right).
\end{split}
\end{equation}
Since the signal model (\ref{E:sig model}) is partially linear, we may concentrate the log-likelihood function by first fixing the parameters $\boldsymbol{\omega}$ and maximizing the likelihood function w.r.t. the linear parameters $\boldsymbol{\phi}$, and then substitute the result back into the likelihood function \cite{Stoica1995}. For a fix value $\boldsymbol{\omega}^\ast$, the solution to (\ref{E:MLE}) is given by the weighted least squares solution \cite{Kay}. That is,
\begin{equation}\label{E:LS}
\begin{split}
\hat{\boldsymbol{\phi}}(\boldsymbol{\omega}^\ast)&=\argmax_{\boldsymbol{\phi}}\left(L(\boldsymbol{\omega}^\ast,\boldsymbol{\phi})\right)\\
&=(\mathbf{H}^\top\mathbf{Q}^{-1}\mathbf{H})^{-1}\mathbf{H}^\top\mathbf{Q}^{-1}(\mathbf{y}-\mathbf{h}(\boldsymbol{\omega}^\ast)).
\end{split}
\end{equation}
Substituting (\ref{E:LS}) back into (\ref{E:MLE}) yields the concentrated maximum likelihood estimator
\begin{equation}\label{E:CMLE}
\hat{\boldsymbol{\omega}}=\argmax_{\boldsymbol{\omega}}\left(L_c(\boldsymbol{\omega})\right),
\end{equation}
where
\begin{equation}\label{E:CLLH}
L_c(\boldsymbol{\omega})\equiv L(\boldsymbol{\omega},\hat{\boldsymbol{\phi}}(\boldsymbol{\omega}))=-\frac{1}{2}\|\mathbf{y}-\mathbf{h}(\boldsymbol{\omega})\|^2_\mathbf{P}+c
\end{equation}
and
\begin{equation}\label{E:P}
  \mathbf{P}=\mathbf{Q}^{-1}-\mathbf{Q}^{-1}\mathbf{H}(\mathbf{H}^\top\mathbf{Q}^{-1}\mathbf{H})^{-1}\mathbf{H}^\top\mathbf{Q}^{-1}.
\end{equation}
%
%\begin{equation}\label{E:P}
%  \mathbf{P}=\mathbf{Q}^{-1}(\mathbf{I}_{3N_{\mbox{\tiny{tot}}}}-\mathbf{H}(\mathbf{H}^\top\mathbf{Q}^{-1}\mathbf{H})^{-1}\mathbf{H}^\top\mathbf{Q}^{-1}).
%\end{equation}
%
Thus, maximizing the concentrated likelihood problem is the equivalent to solving a nonlinear least squares problem. Once the angular velocity estimate $\hat{\boldsymbol{\omega}}$ has been found, then the maximum likelihood estimate $\hat{\boldsymbol{\phi}}$ can be calculated via (\ref{E:LS}).

As the error characteristics of ultralow-cost inertial sensors generally vary with motion dynamics and environmental factors such as temperature \cite{El-Diasty2007}, the assumption that the error covariance matrix $\mathbf{Q}$ is known may not be realistic in all situations. The covariance matrix may then be parameterized and included in the likelihood function, see e.g., \cite{Tao2011}. A few other approaches for estimating the measurement error statistics in inertial sensors arrays are discussed in \cite{Waegli2010}. However, throughout the paper we will assume $\mathbf{Q}$ to be known with sufficient accuracy.

Note that the proposed estimator, i.e., (\ref{E:LS})-(\ref{E:P}), can also be derived without the assumption of the Gaussian distributed measurement error using the least squares framework. The assumption was introduced to provide stringency with the assumptions used to derive the CRB presented in Section~\ref{S:CRB}.

\subsection{Maximizing the concentrated likelihood function}\label{S:Gauss Newton}
The concentrated maximum likelihood estimation problem in (\ref{E:CMLE}) may be solved numerically using the Gauss-Newton algorithm \cite{Kay}. That is, $\hat{\boldsymbol{\omega}}$ can be iteratively calculated as
\begin{equation}\label{E:Iteration}
\hat{\boldsymbol{\omega}}_{k+1}=\hat{\boldsymbol{\omega}}_k+(\mathbf{J}^\top_{h}\mathbf{P}\mathbf{J}_{h})^{-1}\mathbf{J}^\top_{h}\mathbf{P}\bigl(\mathbf{y}-\mathbf{h}(\hat{\boldsymbol{\omega}}_k)\bigr),
\end{equation}
where $k$ denotes the iteration index and $\hat{\boldsymbol{\omega}}_{0}$ is an initial estimate of the angular velocity. A good initial estimate $\hat{\boldsymbol{\omega}}_0$ to the Gauss-Newton algorithm is given by the weighted least squares estimates of the angular velocity calculated from the gyroscope readings, i.e., $\hat{\boldsymbol{\omega}}_0=((\mathbf{1}_{\Nw}^\top\otimes\mathbf{I}_3)\mathbf{Q}^{-1}_\omega(\mathbf{1}_{\Nw}\otimes\mathbf{I}_3))^{-1}(\mathbf{1}_{\Nw}^\top\otimes\mathbf{I}_3)\mathbf{Q}^{-1}_\omega\yw$, where $\mathbf{Q}_\omega$ denotes the covariance matrix of the gyroscopes measurement error $\mathbf{n}_\omega$. The Jacobian $\mathbf{J}_{h}$ of $\mathbf{h}(\boldsymbol{\omega})$ is given by
\begin{equation}\label{E:Jacobian}
  \mathbf{J}_{h} =\begin{bmatrix}\mathbf{A}(\boldsymbol{\omega},\mathbf{r}_1)^\top & \hdots & \mathbf{A}(\boldsymbol{\omega},\mathbf{r}_{\Ns})^\top & \mathbf{1}^\top_{\Nw}\otimes\mathbf{I}_3 \\
  \end{bmatrix}^\top,
\end{equation}
where $\mathbf{A}(\mathbf{u},\mathbf{v})=\mathbf{\Omega}^\top_\mathbf{u}\mathbf{\Omega}_\mathbf{v}+\mathbf{\Omega}_{\mathbf{v}\times\mathbf{u}}$.
%
%\begin{equation}\label{E:gradient g}
%   \mathbf{A}(\mathbf{u},\mathbf{v})=\mathbf{\Omega}^\top_\mathbf{u}\mathbf{\Omega}_\mathbf{v}+\mathbf{\Omega}_{(\mathbf{v}\times\mathbf{u})}.
%\end{equation}
%

The performance of the proposed measurement fusion method is evaluated in Section~\ref{S:exp}, where it is compared to the CRB derived in Section~\ref{S:CRB}. Next, we will describe how the dynamic range of the angular velocity measurements of the array can be extended beyond the dynamic range of the gyroscopes used in the array.

\section{Angular Velocity Measurement Range Extension}
The measurement range of contemporary ultralow-cost gyroscopes is generally limited to some thousands of degrees per second, whereas accelerometers with measurement ranges of thousands of meters per second square are widely available. Thus, when designing sensor arrays to be used in high dynamic applications such as biomedical tracking systems for sport activities where the angular velocity may exceed thousands of degrees per second, see e.g.,~\cite{Lapinski2009,Camarillo2013}, it may be difficult to find gyroscopes with a sufficient dynamic range. Consequently, the gyroscopes in the array sometimes get saturated. However, the proposed measurement fusion method can still be used to estimate the angular velocity by simply removing the saturated gyroscope readings from the measurement model and adapting the dimensions of the measurement error covariance matrix accordingly\footnote{Theoretically, simply removing the measurements that are believed to be saturated will create a bias in the estimated quantities, however our practical experience shows that this effect is negligible.}. %That is, in the case with saturated gyroscopes the measurement model to be used in the measurement fusion is given by
%
%\begin{equation}\label{E:sig model sat}
%\begin{split}
%\mathbf{y}_\sat=\mathbf{T}_\sat(\boldsymbol{\omega},\gamma)\bigl(\mathbf{h}(\boldsymbol{\omega})+\mathbf{H}\,\boldsymbol{\phi}+\mathbf{n}\bigr).
%\end{split}
%\end{equation}
%%
%Here $\mathbf{T}_\sat(\boldsymbol{\omega},\gamma)$ denotes a selection matrix that removes the rows corresponding to the gyroscopes that are saturated. That is,
%
%\begin{equation}\label{E:Tsat}
%  \mathbf{T}_\sat(\boldsymbol{\omega},\gamma)=\mathbf{I}_{3\,\Ns}\oplus \mathbf{E}(\boldsymbol{\omega},\gamma) \oplus \cdots \oplus \mathbf{E}(\boldsymbol{\omega},\gamma),
%\end{equation}
%%
%where $\mathbf{E}(\boldsymbol{\omega},\gamma)=\delta(\omega_x,\gamma)\oplus \delta(\omega_y,\gamma) \oplus \delta(\omega_z,\gamma)$ and
%
%\begin{equation}\label{E:E}
%  \delta(z,\gamma)=\left\{
%                                             \begin{array}{ll}
%                                               1, & |z|\leq\gamma \\
%                                               0, & |z|>\gamma
%                                             \end{array}
%                                           \right..
%\end{equation}
%%
%Here $\oplus$ and $\omega_j$ denote the direct sum operator and the element of the angular velocity vector corresponding to rotations along the $j$:th coordinate axis, respectively. Further, $\gamma$ denotes the saturation level of the gyroscopes in the array.

\begin{figure}[t!]
  \centering
  \psfrag{time}[c][c][0.7][0]{time $[s]$}
      \psfrag{saturation level}[c][c][0.65][0]{Gyroscope saturation level}
   \subfigure[Overlaid measurements form all of the accelerometers in the array. The differences between the measurements are caused by the spatial separation of the sensors and can be used to estimate the angular velocity and angular acceleration of the array.]{%
     \psfrag{title}[c][c][0.75][0]{Accelerometer measurements versus time.}
     \psfrag{ys}[c][c][0.7][0]{$\mathbf{s}$ $[m/s^2]$}
       \psfrag{x-axis}[l][l][0.6][0]{x-axes}
    \psfrag{y-axis}[l][l][0.6][0]{y-axes}
      \psfrag{z-axis}[l][l][0.6][0]{z-axes}
            \label{F:Specific force measurements}
            \includegraphics[width=0.95\columnwidth]{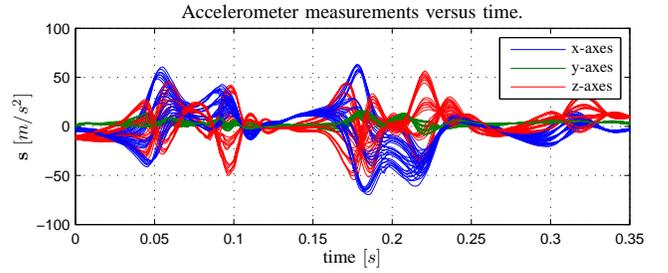}
        }\\%
         \subfigure[Overlaid measurements form all of the gyroscopes in the array. The gyroscopes saturates at 2000 $^\circ/s$.]{%
           \psfrag{title}[c][c][0.75][0]{Gyroscope measurements versus time.}
           \psfrag{ys}[c][c][0.7][0]{$\boldsymbol{\omega}$ $[^\circ/s]$}
                 \psfrag{x-axis}[l][l][0.6][0]{x-axes}
    \psfrag{y-axis}[l][l][0.6][0]{y-axes}
      \psfrag{z-axis}[l][l][0.6][0]{z-axes}
            \label{F:Angular velocity measurements}
            \includegraphics[width=0.95\columnwidth]{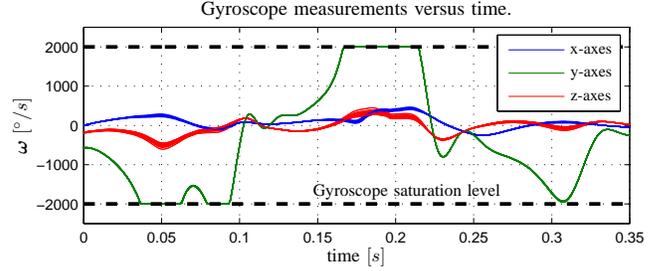}
        }\\
                 \subfigure[Angular velocity estimated using the proposed method. The angular velocity can be estimated even though the gyroscopes are saturated.]{%
        \psfrag{title}[c][c][0.75][0]{Estimated angular velocity versus time.}
        \psfrag{ys}[c][c][0.7][0]{$\boldsymbol{\omega}$ $[^\circ/s]$}
          \psfrag{x-axis}[l][l][0.6][0]{x-axis}
    \psfrag{y-axis}[l][l][0.6][0]{y-axis}
      \psfrag{z-axis}[l][l][0.6][0]{z-axis}
            \label{F:Estimated angular velocity}
            \includegraphics[width=0.95\columnwidth]{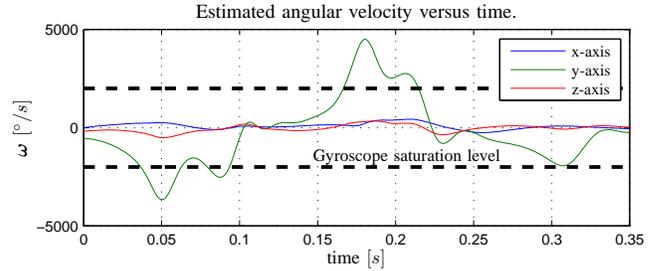}
       }%
 \caption{An illustration of the angular velocity measurement range extension using the proposed method. The measurements were recorded while twisting the array in Fig.~\ref{F:array} by hand. The differences between the accelerometer measurements, see Fig. \ref{F:Specific force measurements}, are used to estimate the angular velocity,  see Fig. \ref{F:Estimated angular velocity},  even when the gyroscopes are saturated, see Fig. \ref{F:Angular velocity measurements}.}\label{F:example dynamic range extension}
\end{figure}

When all of the gyroscopes are saturated, then the measurement model (\ref{E:sig model}) becomes the equivalent to the model in (\ref{E:sig model accelerometer}), for which the angular velocity is not identifiable since $\mathbf{h}_s(\boldsymbol{\omega})=\mathbf{h}_s(-\boldsymbol{\omega})$. However, since the sign of the elements in the angular velocity vector can be deduced from the saturated gyroscopes, the angular velocity is still identifiable. %In practice the sign information can be introduced by initializing the iteration~\eqref{E:Iteration} with the sign obtained from the saturated gyroscopes.
% However, if the saturation levels of the gyroscopes are much larger than the standard deviation of the measurement error, then for all practical applications the sign of the angular velocity can be assumed given by the saturated gyroscope readings.
In Fig.~\ref{F:example dynamic range extension} an example of the angular velocity measurement range extension is shown, when applied to data collected by the array in Fig.~\ref{F:array}.

When the gyroscopes are saturated, finding a good initialization of the Gauss-Newton algorithm becomes trickier and the convergence of the algorithm can be slow. In applications where the inertial sensor array is used on a regular basis to measure the specific force and angular velocity, an initial estimate may be found by predicting the angular velocity from the at pervious time instant estimated angular velocity and angular acceleration \cite{Parsa2007}. Another method to find an initial estimate when the gyroscopes are saturated and a 3D array is used is through the angular acceleration tensor method that is summarized in the Appendix.

The performance of the proposed measurement fusion method, for the case when the gyroscopes are saturated, is evaluated in Section~\ref{S:exp}. The performance is compared to the CRB derived in the next section.

\section{Cram\'{e}r-Rao Bound}\label{S:CRB}
Since the measurement error in the signal model is assumed to be Gaussian distributed, a general expression for the CRB for the measurement fusion problem at hand is straight forward to derive starting from the results in \cite[p.49]{Kay}. Assuming the following parameter order $\boldsymbol{\theta}=\begin{bmatrix}\boldsymbol{\omega}^\top &\boldsymbol{\phi}^\top\end{bmatrix}^\top\equiv\begin{bmatrix}\boldsymbol{\omega}^\top &\dot{\boldsymbol{\omega}}^\top & \mathbf{s}^\top\end{bmatrix}^\top$, the CRB becomes
\begin{equation}\label{E:CRB}
  \mbox{Cov}(\boldsymbol{\theta})\succeq \I(\boldsymbol{\theta})^{-1},
\end{equation}
and the Fisher information matrix $\mathcal{I}(\boldsymbol{\theta})$ is given by
\begin{equation}\label{E:J}
\I(\boldsymbol{\theta})=\mathbf{\Phi}^\top \mathbf{Q}^{-1}\mathbf{\Phi},
\end{equation}
where $\mathbf{\Phi}=\begin{bmatrix}\mathbf{J}_h & \mathbf{H} \end{bmatrix}$. Here the notation  $\mathbf{B} \succeq  \mathbf{A}$ implies that $(\mathbf{B}-\mathbf{A})$ is a positive semi-definite matrix. Since the measurement model is linear in $\boldsymbol{\phi}$, the Fisher information matrix will not depend on the specific force and the angular acceleration. To gain a deeper understanding of the estimation problem at hand, we will study the CRB for a special set of array configurations next.

\subsection{Square arrays with uncorrelated measurement errors}\label{S:CaseI}
Consider the case where the following three conditions apply: (i) the accelerometers are mounted in a planar square grid with a spacing $\alpha$ in between each sensor and the origin of the array coordinate system is defined at the center of the grid; (ii) the measurement errors are uncorrelated;  and (iii) all of the accelerometers and gyroscopes have the same error variance. Then $\mathbf{Q}=(\sigma^2_s\mathbf{I}_{3 \Ns})\oplus(\sigma^2_\omega\mathbf{I}_{3 \Nw})$, where $\oplus$ denotes the matrix direct sum. Further, $\sigma^2_s$ and $\sigma^2_\omega$ denote the measurement error variance of the accelerometers and gyroscopes, respectively. Moreover, thanks to the symmetry of the array $\sum_{i=1}^{\Ns}\mathbf{r}_i=\mathbf{0}_{3,1}$. The Fisher information matrix then takes the following form
\begin{equation}\label{E:J special case}
\I(\boldsymbol{\theta})=
\begin{bmatrix}
\I_{11} & \I_{12} & \mathbf{0}_{3,3} \\
\I_{12}^\top  &\I_{22} & \mathbf{0}_{3,3} \\
\mathbf{0}_{3,3} & \mathbf{0}_{3,3} &  \frac{\Ns}{\sigma^2_s}\mathbf{I}_3
\end{bmatrix},
\end{equation}
where
\begin{eqnarray}
 %\I_{11}&=&\frac{\Nw}{\sigma^2_\omega}\mathbf{E}(\boldsymbol{\omega},\gamma)+\frac{1}{\sigma^2_s}\mathbf{J}^\top_{h_s}\mathbf{J}_{h_s}\\
  \I_{11}&=&\frac{\Nw}{\sigma^2_\omega}\mathbf{I}_3+\frac{1}{\sigma^2_s}\mathbf{J}^\top_{h_s}\mathbf{J}_{h_s}\\
  \I_{12} &=& \frac{1}{\sigma^2_s}\mathbf{J}^\top_{h_s}\mathbf{G}\\
  \I_{22} &=&  \frac{1}{\sigma^2_s}\mathbf{G}^\top\mathbf{G}.
\end{eqnarray}
Here $\mathbf{J}_{h_s}$ denotes the Jacobian of $\mathbf{h}_s(\boldsymbol{\omega})$. From the Fisher information matrix in (\ref{E:J special case}) we can see that for the considered type of array, the CRB for the specific force is given by $\mbox{Cov}(\hat{\mathbf{s}})\succeq \I^{-1}_s$, where $\I^{-1}_s=\frac{\sigma^2_s}{\Ns}\mathbf{I}_3$. That is, the accuracy with which the specific force can be estimated is inversely proportional to the number of accelerometer triads in the array, and independent of the angular velocity and geometry of the array. This is quite intuitively since for a symmetric array with a set of accelerometer triads having the same measurement error characteristics (measurement error covariance), the maximum likelihood estimator for the specific force is given by the arithmetic mean of the accelerometer measurements.

Using the Schur complement, the CRB for the angular velocity can be found to be
\begin{equation}\label{E:CRB omega}
  \mbox{Cov}(\hat{\boldsymbol{\omega}})\succeq \I^{-1}_\omega
\end{equation}
where
\begin{equation}\label{E:J omega}
\begin{split}
 \I_\omega&=\I_{11}-\I_{12}\I^{-1}_{22}\I^\top_{12}\\
 &=\frac{\Nw}{\sigma^2_\omega}\mathbf{I}_3+\frac{1}{\sigma^2_s}\bigl(\mathbf{\Gamma}_{11}-\mathbf{\Gamma}_{12}\mathbf{\Gamma}^{-1}_{22}\mathbf{\Gamma}^\top_{12}\bigr),
\end{split}
\end{equation}
and
\begin{eqnarray}\label{E:Gamma1}
  \mathbf{\Gamma}_{11} &=& \mathbf{J}^\top_{h_s}\mathbf{J}_{h_s}=\sum_{i=1}^{\Ns}\mathbf{A}(\boldsymbol{\omega},\mathbf{r}_i)^\top\mathbf{A}(\boldsymbol{\omega},\mathbf{r}_i)\\
  \mathbf{\Gamma}_{12} &=& \mathbf{J}^\top_{h_s}\mathbf{G}=-\sum_{i=1}^{\Ns}\mathbf{A}(\boldsymbol{\omega},\mathbf{r}_i)^\top\mathbf{\Omega}_{\mathbf{r}_i}\\
    \mathbf{\Gamma}_{22} &=& \mathbf{G}^\top\mathbf{G}=\sum_{i=1}^{\Ns}\mathbf{\Omega}_{\mathbf{r}_i}^\top\mathbf{\Omega}_{\mathbf{r}_i}.\label{E:Gamma3}
\end{eqnarray}
Next, using the symmetric properties of the considered array geometry and performing, tedious but straight forward, calculations give that

\begin{equation}\label{E:J omega2}
\begin{split}
 %\I_\omega&=\frac{\Nw}{\sigma^2_\omega}\mathbf{E}(\boldsymbol{\omega},\gamma)\\&+\frac{\alpha^2(\Ns^2-\Ns)}{6\,\sigma^2_s}  %
 \I_\omega&=\frac{\Nw}{\sigma^2_\omega}\mathbf{I}_3\\&+\frac{\alpha^2(\Ns^2-\Ns)}{6\,\sigma^2_s}
 \begin{bmatrix} 2\omega^2_x+\omega^2_y & \omega_x\omega_y & 2\omega_x\omega_z\\
 \omega_x\omega_y & 2\omega^2_y+\omega^2_x & 2\omega_y\omega_z\\
 2\omega_x\omega_z & 2\omega_y\omega_z &  4\omega^2_z\end{bmatrix}.
 \end{split}
\end{equation}
Note that the assumption about the accelerometers being mounted in a planar square grid implies that $\Ns\in\{4,9,16,25,\ldots\}$. From the expression for the Fisher information matrix in (\ref{E:J omega2}), we can see that when the array is stationary no rotational information is gained from the accelerometers, and the CRB becomes equivalent to the covariance of the arithmetic mean of the gyroscope measurement errors. When the array starts rotating, the accelerometers will start to provide rotational information and the accuracy with which the angular velocity can be estimated increases proportionally to the squared angular rates. Further, the rotational information gained from the accelerometers scale quadratically with the distance $\alpha$ between the sensors.

When all gyroscopes are saturated, i.e., $|\omega_i|>\gamma,~\forall i\in\{x,y,z\}$, where $(-\gamma,\gamma)$ is the dynamic range of the gyroscopes, no information other than the sign of the angular velocity is provided by the gyroscopes and the angular velocity must be estimated sole from the accelerometers. As the sign information can be seen as a special case of inequality constraints, and inequality constraints generally does not have any effect on the CRB \cite{Gorman1990}, the Fisher information for this case can be found by letting $\sigma^2_\omega\rightarrow\infty$. The inverse of the Fisher information matrix (\ref{E:J omega2}) then takes the following simple form

\begin{equation}\label{E:J inv omega sat}
\begin{split}
 \I_\omega^{-1}&\stackrel{\scriptscriptstyle{(|\omega|>\gamma)}}{=}\frac{6\,\sigma^2_s}{\alpha^2(\Ns^2-\Ns)} \begin{bmatrix} \frac{1}{\omega^2_x+\omega^2_y} & 0 & \bullet\\
 0 & \frac{1}{\omega^2_x+\omega^2_y} & \bullet\\
 \bullet & \bullet & \frac{1}{2\omega^2_z}\end{bmatrix},
 \end{split}
\end{equation}
where the non-zero of the diagonal elements have been left out for brevity. From this, we can see that in an array where the gyroscopes have an infinitely small dynamic range, i.e., the gyroscopes only provides information regarding the sign of the angular velocity, then the covariance of the angular velocity estimates tends toward infinity as the angular velocity goes toward zero. Hence, for so called gyroscope-free IMUs, i.e., an array of only accelerometers, there is no estimator that can provide an unbiased, finite variance, estimate of small angular velocities. %not even if the sign of the angular velocity is known.
Therefore, the practical use of gyroscope-free IMUs in inertial navigation systems for low-dynamical applications is questionable.

The CRB for the angular acceleration $\dot{\boldsymbol{\omega}}$ is given by
\begin{equation}\label{E:CRB omega dot}
\begin{split}
  \mbox{Cov}(\hat{\dot{\boldsymbol{\omega}}})&\succeq \I^{-1}_{\dot{\omega}},
\end{split}
\end{equation}
where
\begin{equation}\label{E:J omega dot}
\begin{split}
  \I_{\dot{\omega}}&=\I_{22}-\I_{12}^\top\I^{-1}_{11}\I_{12}\\
  &=\frac{1}{\sigma^2_s}\mathbf{\Gamma}_{22}-\frac{1}{\sigma^4_s}\mathbf{\Gamma}^\top_{12}\Bigl(\frac{\Nw}{\sigma^2_\omega}\mathbf{I}_3+\frac{1}{\sigma^2_s}\mathbf{\Gamma}_{11}\Bigr)^{-1}\mathbf{\Gamma}_{12}.
 \end{split}
\end{equation}
From (\ref{E:J omega dot}) we can note that the accuracy with which the angular acceleration $\dot{\boldsymbol{\omega}}$ can be estimated, somewhat surprisingly, decreases with an increasing angular velocity. The best accuracy is achieved when the array is in linear motion ($\boldsymbol{\omega}=\mathbf{0}$) and is given by
\begin{equation}\label{E:CRB omega dot lower}
  \I^{-1}_{\dot{\omega}} \stackrel{\scriptscriptstyle{(\omega=0)}}{=}\sigma^2_s\mathbf{\Gamma}_{22}^{-1}=\frac{12\,\sigma^2_s}{\alpha^2(\Ns^2-\Ns)} \begin{bmatrix} 1 & 0 & 0\\
 0 & 1 & 0\\
 0 & 0 & \frac{1}{2}\end{bmatrix}.
\end{equation}

\section{Simulation and Experiments}\label{S:exp}
In this section the performance of the proposed maximum likelihood measurement fusion method will be evaluated through simulations and real-world experiments. The results will be compared to the derived CRB.

\subsection{Simulations}

\begin{figure}[t!]
  \centering
   \psfrag{a}[c][c][0.6][0]{accelerometer}
   \psfrag{t}[c][c][0.6][0]{triad}
   \psfrag{z}[l][l][0.6][0]{z-axis}
   \psfrag{y}[l][l][0.6][0]{y-axis}
   \psfrag{x}[l][l][0.6][0]{x-axis}
   \psfrag{e}[c][c][0.5][0]{1[$cm$]}
   \psfrag{e1}[c][c][0.5][45]{1[$cm$]}
   \psfrag{e2}[c][c][0.5][90]{1[$cm$]}
   \psfrag{i}[c][c][0.6][0]{in-plan rotation}
   \psfrag{o}[c][c][0.6][0]{out-of-plan rotation}
    \subfigure[Planar geometry]{%
            \label{F:planar geometry}
            \includegraphics[width=0.4\columnwidth]{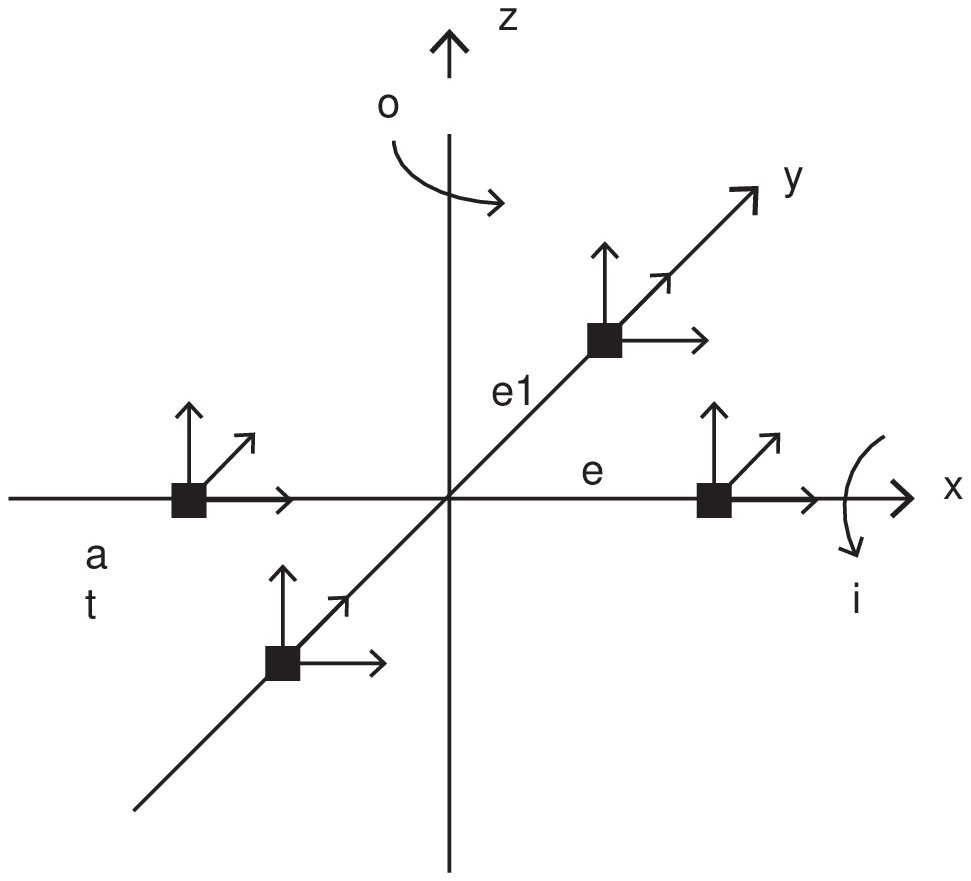}}\hspace{0.1\columnwidth}
                \subfigure[Non-planar geometry]{%
            \label{F:non planar geometry}
            \includegraphics[width=0.4\columnwidth]{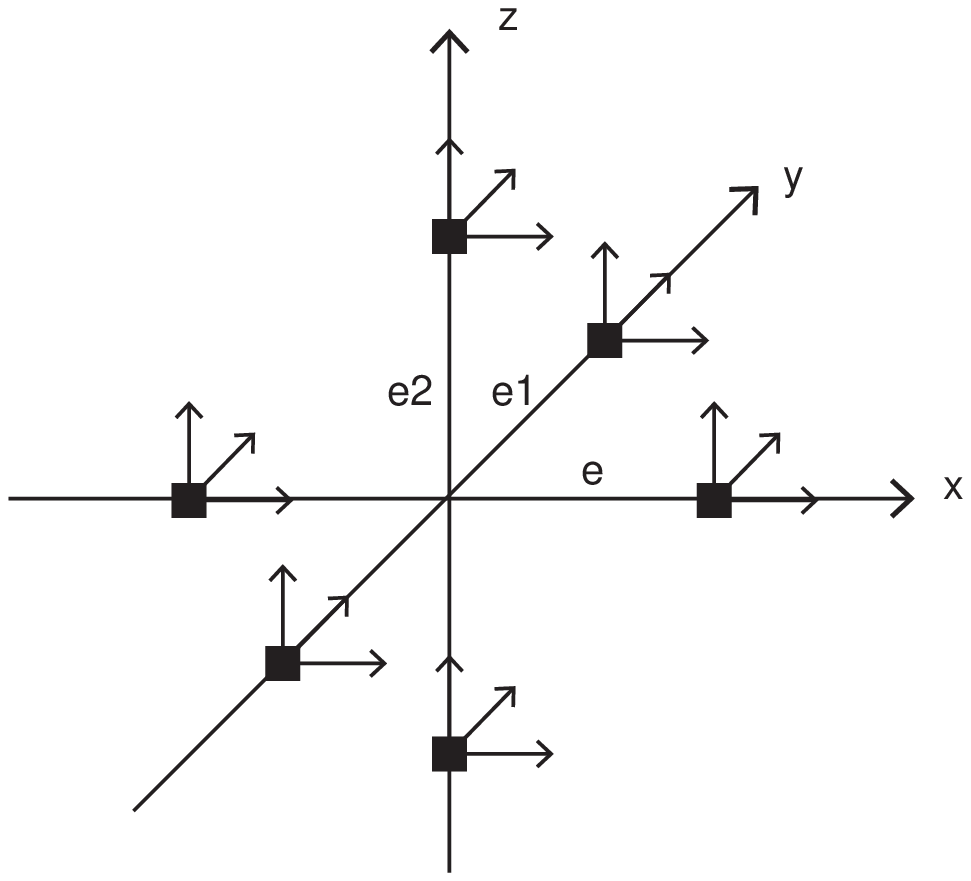}}
  \caption{Geometries of the accelerometer triads used in the Monte Carlo simulations.}\label{F:array_illustration}
\end{figure}

To evaluate the accuracy of the estimator we have conducted two Monte-Carlo simulations, where each Monte-Carlo simulation runs the proposed measurement fusion method on $10^5$ data realizations. The array considered in the simulations consists of four accelerometer triads and four gyroscope triads, where the accelerometer triads are mounted as illustrated in Fig. \ref{F:planar geometry}. Note that all of the arrays that fulfill the assumptions made in Section~\ref{S:CaseI} can be transformed into an equivalent array with a geometry as in Fig. \ref{F:planar geometry} plus possible additional sensors located at the origin. The measurement errors of the accelerometers and the gyroscopes were assumed to be uncorrelated and to have the standard deviation  $\sigma^2_s=0.01$ \,[$m/s^2$] and $\sigma_\omega=1$\, [$^\circ\!/s^2$], respectively. Further, the gyroscopes were assumed to saturate at 2000\,[$^\circ\!/s^2$]. These parameter values were selected to reflect the performance that can be expected from ultralow-cost inertial sensors during high dynamic operations when scale-factor, g-sensitivity, and cross-axis sensitivity errors become the main error sources. The small value selected for the accelerometer measurement error variance is motivated by the fact that the accelerometers in the array can easily be calibrated using e.g., the method in \cite{NSH2014}. A calibration of the gyroscopes is complicated and requires a rotational rig. If the accelerometers where uncalibrated, then the standard deviation of the accelerometer measurement errors would be a magnitude higher approximately; refer to \cite{NSH2014} for typical error figures before and after sensor calibrations.

\begin{figure}[t!]
  \centering
  \psfrag{ylabel}[c][c][0.7][0]{RMSE [$^\circ\!/s$]}
  \psfrag{xlabel}[c][c][0.7][0]{$\|\boldsymbol{\omega}\|$ [$^\circ\!/s$]}
  \psfrag{base level}[c][c][0.8][0]{Base level: $\sigma_\omega/\sqrt{4}$}
  \psfrag{x}[l][l][0.8][0]{$\omega_x$ Maximum likelihood}
  \psfrag{y}[l][l][0.8][0]{$\omega_y$ Maximum likelihood}
  \psfrag{z}[l][l][0.8][0]{$\omega_z$ Maximum likelihood}
  \psfrag{Average gyroscopes}[l][l][0.8][0]{Average gyroscopes}
  \psfrag{CRB}[l][l][0.8][0]{$\sqrt{\mbox{CRB}}$}
  \psfrag{CRB saturated gyroscope xxxx}[l][l][0.8][0]{$\sqrt{\mbox{CRB}}$ saturated gyroscope}
   \subfigure[Array rotating around x-axis, i.e., $\boldsymbol{\omega}=\|\boldsymbol{\omega}\|{[}\,1\,0\,0\,{]}^\top$.]{%
     \psfrag{title}[c][c][0.75][0]{Angular velocity estimation (in-plan rotation).}
     \psfrag{Gyroscope}[l][l][0.8][0]{Gyroscope}
     \psfrag{saturation}[l][l][0.8][0]{x-axis satu-}
     \psfrag{level}[l][l][0.8][0]{ration level}
            \label{F:MC_inplane}
            \includegraphics[width=\columnwidth]{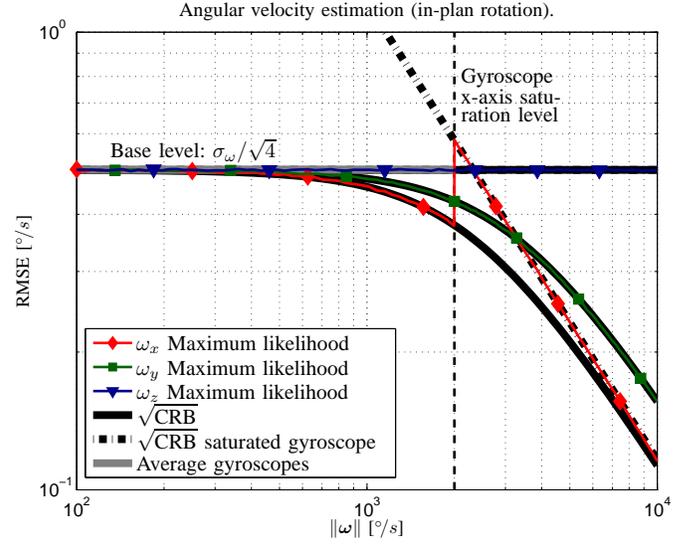}
        }\\%
         \subfigure[Array rotating around z-axis, i.e., $\boldsymbol{\omega}=\|\boldsymbol{\omega}\|{[}\,0\,0\,1\,{]}^\top$.]{%
           \psfrag{title}[c][c][0.75][0]{Angular velocity estimation (out-of-plan rotation).}
           \psfrag{Gyroscope}[l][l][0.8][0]{Gyroscope}
           \psfrag{saturation}[l][l][0.8][0]{z-axis satu-}
           \psfrag{level}[l][l][0.8][0]{ration level}
            \label{F:MC_outplane}
            \includegraphics[width=\columnwidth]{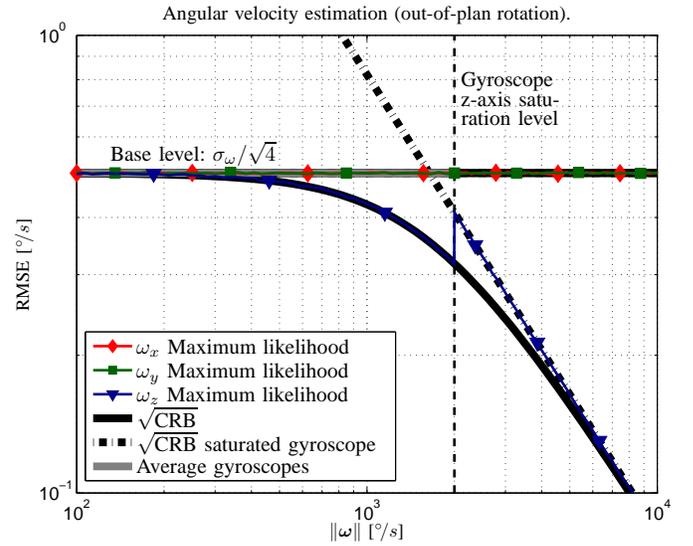}
        }%\\%
%                 \subfigure[Array rotating around all axis, i.e., $\boldsymbol{\omega}=\|\boldsymbol{\omega}\|/\sqrt{3}\,{[}\,1\,1\,1\,{]}^\top$.]{%
%        \psfrag{title}[c][c][0.75][0]{Angular velocity estimation (all axis rotation).}
%            \label{F:MC allplane}
%            \includegraphics[width=0.95\columnwidth]{all_velocity.eps}
%       }%
 \caption{The angular velocity estimation RMSE of the proposed measurement fusion method for a planar array with the geometry illustrated in Fig.~\ref{F:planar geometry}. The estimation accuracy obtained by simply averaging the gyroscope measurements is given by the gray horizontal line.}\label{F:MC_angular_velocity}
%  \end{subfigure}
\end{figure}

The root mean square errors (RMSE) calculated from the Monte-Carlo simulations, together with the corresponding square root of the CRBs, are shown in Fig.~\ref{F:MC_angular_velocity}. In Fig.~\ref{F:MC_inplane} the result when the array is rotated around the x-axis, i.e., an in-plan rotation, is shown. Clearly, no information about the angular rate around the z-axis (pointing out of the plan) is gained from the accelerometers, whereas the accelerometer measurements contribute to the estimation of the angular rates around the x-axis and y-axis. In Fig.~\ref{F:MC_outplane} the result when the array is rotated around the z-axis, i.e., an out-of-plan rotation, is shown. In this case no information about the rotations around the x-axis and y-axis are gained from the accelerometers; only information about the rotation around the z-axis is gained. The estimation accuracy achieved by simply averaging the gyroscope measurements is illustrated by the grey horizontal line in the figures.

\begin{figure}[t!]
  \centering
  \psfrag{ylabel}[c][c][0.7][0]{RMSE [$^\circ\!/s$]}
  \psfrag{xlabel}[c][c][0.7][0]{$\|\boldsymbol{\omega}\|$ [$^\circ\!/s$]}
  \psfrag{Gyroscope}[l][l][0.8][0]{Gyroscope}
  \psfrag{saturation}[l][l][0.8][0]{saturation}
  \psfrag{level}[l][l][0.8][0]{levels}
  \psfrag{base level}[c][c][0.8][0]{Base level: $\sigma_\omega/\sqrt{6}$}
  \psfrag{x}[l][l][0.8][0]{$\omega_x$ Maximum likelihood}
  \psfrag{y}[l][l][0.8][0]{$\omega_y$ Maximum likelihood}
  \psfrag{z}[l][l][0.8][0]{$\omega_z$ Maximum likelihood}
  \psfrag{x t}[l][l][0.8][0]{$\omega_x$ Tensor method}
  \psfrag{y t}[l][l][0.8][0]{$\omega_y$ Tensor method}
  \psfrag{z t}[l][l][0.8][0]{$\omega_z$ Tensor method}
  \psfrag{Average gyroscopes}[l][l][0.8][0]{Average gyroscopes}
  \psfrag{CRB}[l][l][0.8][0]{$\sqrt{\mbox{CRB}}$}
  \psfrag{CRB saturated gyroscope xxxx}[l][l][0.8][0]{$\sqrt{\mbox{CRB}}$ saturated gyroscope}
  \psfrag{title}[c][c][0.75][0]{Maximum likelihood versus tensor method.}
  \includegraphics[width=0.97\columnwidth]{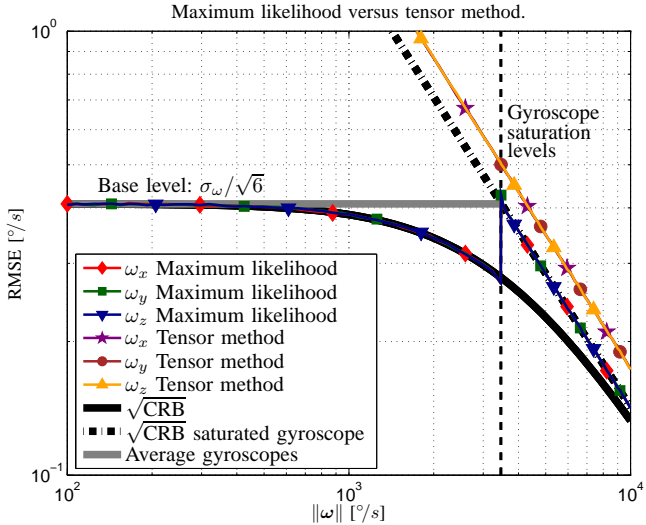}
  \caption{The angular velocity estimation RMSE of the proposed maximum likelihood measurement fusion method and the angular acceleration tensor method. The geometry of the considered array is illustrated in Fig.~\ref{F:non planar geometry} and the array is rotating around all of the axes, i.e., $\boldsymbol{\omega}=\|\boldsymbol{\omega}\|/\sqrt{3}\,{[}\,1\,1\,1\,{]}^\top$. The estimation accuracy obtained by simply averaging the gyroscope measurements is given by the gray horizontal line.}\label{F:MC_allcube}
\end{figure}

To compare the proposed measurement fusion method with the in gyroscope-free IMUs commonly used angular acceleration tensor method, a Monte Carlo simulation using the array geometry illustrated in Fig.~\ref{F:non planar geometry} was also conducted; the angular acceleration tensor method is summarized in the Appendix. The non-planar array geometry is needed for the tensor method to work. The result of the simulation is shown in Fig.~\ref{F:MC_allcube}, from which it can be seen that the tensor method does not achieve the CRB and is outperformed by the proposed method. Note that the tensor method only uses the measurements from the accelerometers and the sign information from the gyroscopes, and therefore the behavior of the two methods should only be compared for the part of the simulation where all of the gyroscopes are saturated, i.e., above $2000\circ/s$.

%\addtolength{\subfigcapskip}{-2.5mm}

\subsection{Experiments}
\begin{figure}[t!]
  \centering
    \psfrag{Gyroscope}[l][l][0.7][0]{Gyroscope}
  \psfrag{saturation}[l][l][0.7][0]{saturation}
  \psfrag{level}[l][l][0.7][0]{level}
  \psfrag{base level}[c][c][0.7][0]{Base level: $\sigma_\omega/\sqrt{32}$}
  \psfrag{experimentaaa}[c][c][0.65][0]{Experiment}
  \psfrag{simulationaaa}[c][c][0.65][0]{Simulation}
  \psfrag{x}[c][c][0.7][0]{$\|\boldsymbol{\omega}\|$ [$^\circ\!/s$]}
  \psfrag{y}[c][c][0.7][0]{RMSE [$^\circ\!/s$]}
  \psfrag{title}[c][c][0.75][0]{Experimental calculated angular speed accuracy.}
  \includegraphics[width=0.97\columnwidth]{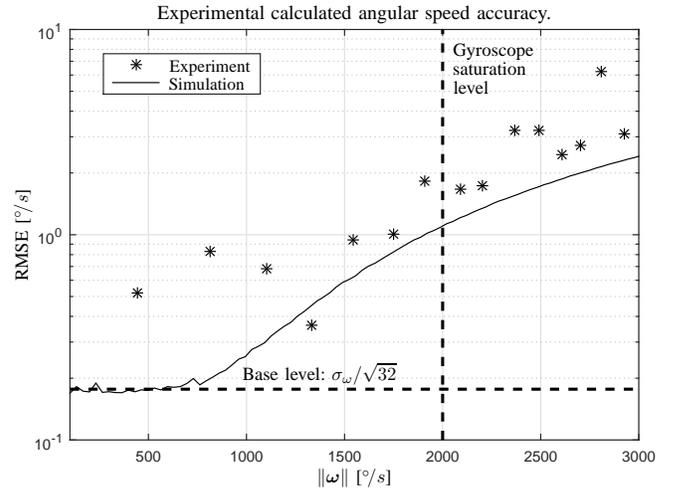}
  \caption{RMSE of the from real-world data calculated angular speed. Also shown, the RMSE of the estimated angular speed when calculated from simulated data that includes random sensor location errors.}\label{F:Expriement results}
\end{figure}

To evaluate the proposed method with real-world data the in-house designed inertial sensor array, shown in Fig.~\ref{F:array}, was mounted in a mechanical rotation rig and data was recorded at different angular speeds. At each angular speed, data corresponding to 100 time instants were recorded. The data was processed using the proposed maximum likelihood method with the same noise variance settings as used in the simulations and the RMSE of the estimated angular speed was calculated; the angular speed and not the angular velocity as an evaluation criterion was chosen because of the practical problem of accurately aligning the coordinate axes of the array with the rotation axis of the rotation rig. The result is displayed in Fig.~\ref{F:Expriement results},  and it shows that the proposed method works but it does not achieve the accuracy predicted by the theoretical model. Several factors such as sensor scale factor errors, misalignment between the sample instances of the sensors, and uncertainties in the location of the sensing elements, have been identified as likely sources for the discrepancy. For illustrational purposes the result of a Monte-Carlo simulation resembling the experimental setup, and where random errors with a standard deviation of 0.1\,mm were added to the sensor locations, is also shown in Fig.~\ref{F:Expriement results}.\footnote{Placement errors in the order of 0.1\,mm are likely to arise in the PCB manufacturing and population process.} More experimental results can be found in \cite{SNH2014b}, where the performance of an earlier revision of the array when used in a foot-mounted inertial navigation system were evaluated.

\section{Summary, Conclusions, and Future Research}
Approximately 5 billion smartphones were sold in the five year span 2011-2015, of which most where equipped with some type of inertial motion sensors. Thus, the smartphone industry has become a driving force in the development of ultralow-cost inertial sensors, and six-degree-of-freedom inertial sensor assemblies are sold to large-volume costumers for less than a dollar. Unfortunately, these ultralow-cost sensors do not yet meet the needs of more demanding applications like inertial navigation and biomedical motion tracking systems. However, by adapting a wisdom of the crowd's thinking and design arrays consisting of hundreds of sensing elements, one can capitalize on the decreasing cost, size, and power-consumption of the sensors to construct virtual high-performance low-cost inertial sensors.  Accordingly, we have proposed a maximum-likelihood method to fuse the measurements in arrays consisting of multiple accelerometer and gyroscope triads. The proposed method has been evaluated through simulations as well as real-world tests with an in-house designed array with 192 sensing elements. The simulation results show that the proposed method attains the CRB and it outperforms the current state-of-the-art method. \emph{Further, compared to the state-of-the-art method the proposed method also works with 2D arrays, which is fundamental for the production of arrays on a single printed circuit board.} Moreover, by utilizing the spatial separation between the accelerometers, the angular velocity measurement range can be extended beyond that of the gyroscopes. The experimental results show that the proposed methods work, but do not achieve the accuracy predicted by the theoretical model. Uncertainties in the position of the sensing elements have been identified as a likely source for the discrepancy. Further research is needed to verify the source of the problem, and new inertial sensor array calibration methods must be developed. Moreover, information fusion methods that also considers the time development of the sensor signals and allows for simultaneous calibration and motion tracking should be investigated. Another open research question is to design a measurement error covariance estimation method that enables automatic weighting of the sensor measurements.

\appendix
The angular acceleration tensor method used in the simulations are here summarized. For a more detailed description see \cite{Parsa2004}. Introducing the angular acceleration tensor $\mathbf{W}=\boldsymbol{\Omega}_{\boldsymbol{\omega}}^2+\boldsymbol{\Omega}_{\dot{\boldsymbol{\omega}}}$  and using the properties of the cross product, the model for the specific force (\ref{E:force eq 2}) can be rewritten as
\begin{equation}\label{E:force tensor}
\mathbf{s}_i=\mathbf{s}+\boldsymbol{\Omega}_{\boldsymbol{\omega}}^2\mathbf{r}_i-\boldsymbol{\Omega}_{\mathbf{r}_i}
\dot{\boldsymbol{\omega}}=\mathbf{s}+\mathbf{W}\,\mathbf{r}_i.
\end{equation}
The output of the $N_s$ accelerometer triads can be described by the following linear matrix equation
\begin{equation}\label{E:Y}
  \mathbf{Y}=\mathbf{X}\mathbf{R}+\mathbf{N}
\end{equation}
where
\begin{eqnarray}
 \mathbf{Y}&=&\left[
                    \begin{array}{ccc}
                      \mathbf{y}_{s_1} & \hdots & \mathbf{y}_{s_{\Ns}} \\
                    \end{array}
                  \right]\quad \mathbf{R}=\left[
                    \begin{array}{ccc}
                      1 & \hdots & 1\\
                      \mathbf{r}_1 & \hdots & \mathbf{r}_{\Ns} \\
                    \end{array}
                  \right]\\
  \mathbf{N}&=&\left[
                    \begin{array}{ccc}
                      \mathbf{n}_{s_1} & \hdots & \mathbf{n}_{s_{\Ns}} \\
                    \end{array}
                  \right]\quad \mathbf{X}=\left[
                    \begin{array}{cc}
                      \mathbf{s} & \mathbf{W}\\
                    \end{array}
                  \right].
\end{eqnarray}
Here $\mathbf{y}_{s_i}$ and $\mathbf{n}_{s_i}$ denote the measurement and measurement error of the $i$:th accelerometer triad. Neglecting the fact that the tensor $\mathbf{W}$ only has six degrees of freedom, the least square estimate of the matrix $\mathbf{X}$ is given by

\begin{equation}
 \widehat{\mathbf{X}}=\mathbf{Y}\mathbf{R}^\top(\mathbf{R}\mathbf{R}^\top)^{-1}.
\end{equation}
Noteworthy is that since $\mathbf{X}$ is a 3-by-4 matrix, the measurements from at least four accelerometer triads are needed. Further, for $\mathbf{R}$ to have a full row rank and the estimation problem to be well defined, the locations of the accelerometer triads must span a three dimensional space. From the estimated tensor $\widehat{\mathbf{W}}$, the angular acceleration can be calculated as
\begin{equation}
  \widehat{\dot{\boldsymbol{\omega}}}=\left[
                                    \begin{array}{ccc}
                                      \widehat{w}_{3,2}-\widehat{w}_{2,3} & \widehat{w}_{1,3}-\widehat{w}_{3,1} & \widehat{w}_{2,1}-\widehat{w}_{1,2} \\
                                    \end{array}
                                  \right]^\top,
\end{equation}
where $\widehat{w}_{i,j}=[\widehat{\mathbf{W}}]_{i,j}$. The angular velocity can, up to a sign ambiguity, be calculated from the left hand side of the equality

\begin{equation}
  \widehat{\boldsymbol{\omega}}\widehat{\boldsymbol{\omega}}^\top=\frac{1}{2}(\widehat{\mathbf{W}}+\widehat{\mathbf{W}}^\top)- \frac{1}{4}\mbox{tr}(\widehat{\mathbf{W}}+\widehat{\mathbf{W}}^\top)\mathbf{I}_3.
\end{equation}
Here $\mbox{tr}(\cdot)$ denotes the trace operator.

\bibliographystyle{IEEEtran}
\bibliography{IEEEabrv,MIMUrefs}

\begin{thebibliography}{10}
\providecommand{\url}[1]{#1}
\csname url@samestyle\endcsname
\providecommand{\newblock}{\relax}
\providecommand{\bibinfo}[2]{#2}
\providecommand{\BIBentrySTDinterwordspacing}{\spaceskip=0pt\relax}
\providecommand{\BIBentryALTinterwordstretchfactor}{4}
\providecommand{\BIBentryALTinterwordspacing}{\spaceskip=\fontdimen2\font plus
\BIBentryALTinterwordstretchfactor\fontdimen3\font minus
  \fontdimen4\font\relax}
\providecommand{\BIBforeignlanguage}[2]{{%
\expandafter\ifx\csname l@#1\endcsname\relax
\typeout{** WARNING: IEEEtran.bst: No hyphenation pattern has been}%
\typeout{** loaded for the language `#1'. Using the pattern for}%
\typeout{** the default language instead.}%
\else
\language=\csname l@#1\endcsname
\fi
#2}}
\providecommand{\BIBdecl}{\relax}
\BIBdecl

\bibitem{Shaeffer2013}
D.~Shaeffer, ``{MEMS} inertial sensors: A tutorial overview,'' \emph{{IEEE}
  Commun. Mag.}, vol.~51, no.~4, pp. 100--109, Apr. 2013.

\bibitem{Perlmutter2012}
M.~Perlmutter and L.~Robin, ``High-performance, low cost inertial {MEMS}: A
  market in motion!'' in \emph{Proc. IEEE/ION Position Location and Navigation
  Symposium (PLANS)}, Myrtle Beach, SC, Apr. 2012, pp. 225--229.

\bibitem{YoleReport}
\BIBentryALTinterwordspacing
G.~Girardin and E.~Mounier, ``6- \& 9-axis sensors consumer inertial combos,''
  Yole D\'{e}veloppement, Tech. Rep., 2014. [Online]. Available:
  \url{http://www.yole.fr/iso\_upload/Samples/Yole\_6\_and\_9-Axis\_Sensors\_Consumer\_Inertial\_Combos.pdf}
\BIBentrySTDinterwordspacing

\bibitem{Bittner2014}
D.~Bittner, J.~Christian, R.~Bishop, and D.~May, ``Fault detection, isolation,
  and recovery techniques for large clusters of inertial measurement units,''
  in \emph{Proc. IEEE/ION Position, Location and Navigation Symposium (PLANS)},
  Monterey, CA, May 2014, pp. 219--229.

\bibitem{Song2015}
J.~W. Song and C.~G. Park, ``Optimal configuration of redundant inertial
  sensors considering lever arm effect,'' \emph{{IEEE} Sensors J.}, vol.~16,
  no.~9, pp. 3171--3180, May 2016.

\bibitem{Waegli2010}
A.~Waegli, J.~Skaloud, S.~Guerrier, M.~Parés, and I.~Colomina, ``Noise
  reduction and estimation in multiple micro-electro-mechanical inertial
  systems,'' \emph{Measurement Science and Technology}, vol.~21, 2010.

\bibitem{SNH2014a}
I.~Skog, J.-O. Nilsson, and P.~H\"{a}ndel, ``An open-source multi inertial
  measurement unit {(MIMU)} platform,'' in \emph{Proc. International Symposium
  on Inertial Sensors and Systems (ISISS)}, Laguna Beach, CA, USA, Feb. 2014.

\bibitem{Krim1996}
H.~Krim and M.~Viberg, ``Two decades of array signal processing research: the
  parametric approach,'' \emph{{IEEE} Signal Process. Mag.}, vol.~13, no.~4,
  pp. 67--94, Jul. 1996.

\bibitem{Stoica1989}
P.~Stoica and A.~Nehorai, ``{MUSIC}, maximum likelihood, and {C}ramer-{R}ao
  bound,'' \emph{{IEEE} Trans. Acoust., Speech, Signal Process.}, vol.~37,
  no.~5, pp. 720--741, May 1989.

\bibitem{Viberg1991}
M.~Viberg and B.~Ottersten, ``Sensor array processing based on subspace
  fitting,'' \emph{{IEEE} Trans. Signal Process.}, vol.~39, no.~5, pp.
  1110--1121, May 1991.

\bibitem{Nehorai1994}
A.~Nehorai and E.~Paldi, ``Vector-sensor array processing for electromagnetic
  source localization,'' \emph{{IEEE} Trans. Signal Process.}, vol.~42, no.~2,
  pp. 376--398, Feb. 1994.

\bibitem{Nehorai1994b}
------, ``Acoustic vector-sensor array processing,'' \emph{{IEEE} Trans. Signal
  Process.}, vol.~42, no.~9, pp. 2481--2491, Sep. 1994.

\bibitem{Abdi2009}
A.~Abdi and H.~Guo, ``Signal correlation modeling in acoustic vector sensor
  arrays,'' \emph{{IEEE} Trans. Signal Process.}, vol.~57, no.~3, pp. 892--903,
  Mar. 2009.

\bibitem{Zou2009}
N.~Zou and A.~Nehorai, ``Circular acoustic vector-sensor array for mode
  beamforming,'' \emph{{IEEE} Trans. Signal Process.}, vol.~57, no.~8, pp.
  3041--3052, Aug. 2009.

\bibitem{Jeremic2000}
A.~Jeremic and A.~Nehorai, ``Landmine detection and localization using chemical
  sensor array processing,'' \emph{{IEEE} Trans. Signal Process.}, vol.~48,
  no.~5, pp. 1295--1305, May 2000.

\bibitem{Nilsson2016}
J.-O. Nilsson and I.~Skog, ``Inertial sensor arrays -- {A} literature review,''
  in \emph{Proc. European Navigation Conference}, Helsinki, Finland, May 2016.

\bibitem{Williams2013}
T.~R. Williams, D.~W. Raboud, and K.~R. Fyfe, ``Minimal spatial accelerometer
  configurations,'' \emph{J. Dyn. Sys., Meas., Control}, vol. 135, p. 021016,
  2013.

\bibitem{Tan2005}
C.-W. Tan and S.~Park, ``Design of accelerometer-based inertial navigation
  systems,'' \emph{{IEEE} Trans. Instrum. Meas.}, vol.~54, no.~6, pp.
  2520--2530, Dec. 2005.

\bibitem{Schuler1967}
A.~R. Schuler, A.~Grammatikos, and K.~A. Fegley, ``Measuring rotational motion
  with linear accelerometers,'' \emph{{IEEE} Trans. Aerosp. Electron. Syst.},
  vol. AES-3, no.~3, pp. 465--472, May 1967.

\bibitem{Schopp2014}
P.~Schopp, H.~Graf, M.~Maurer, M.~Romanovas, L.~Klingbeil, and Y.~Manoli,
  ``Observing relative motion with three accelerometer triads,'' \emph{{IEEE}
  Trans. Instrum. Meas.}, vol.~63, no.~12, pp. 3137--3151, Dec. 2014.

\bibitem{Sukkarieh2000}
S.~Sukkarieh, P.~Gibbens, B.~Grocholsky, K.~Willis, and H.~F. Durrant-Whyte,
  ``A low-cost, redundant inertial measurement unit for unmanned air
  vehicles,'' \emph{Int. J. Robot. Res.}, vol.~19, pp. 1089--1103, 2000.

\bibitem{Chatterjee2015}
G.~Chatterjee, L.~Latorre, F.~Mailly, P.~Nouet, N.~Hachelef, and C.~Oudea,
  ``Smart-mems based inertial measurement units: gyro-free approach to improve
  the grade,'' \emph{Microsystem Technologies}, pp. 1--10, Dec. 2015.

\bibitem{Naseri2014}
H.~Naseri and M.~Homaeinezhad, ``Improving measurement quality of a
  {MEMS}-based gyro-free inertial navigation system,'' \emph{Sensors and
  Actuators A: Physical}, vol. 207, pp. 10--19, Mar. 2014.

\bibitem{He2012}
P.~He and P.~Cardou, ``Estimating the angular velocity from body-fixed
  accelerometers,'' \emph{J. Dyn. Sys., Meas., Control}, vol. 134, no.~6, p.
  061015, 2012.

\bibitem{Madgwick2013}
S.~O. Madgwick, A.~J. Harrison, P.~M. Sharkey, R.~Vaidyanathan, and W.~S.
  Harwin, ``Measuring motion with kinematically redundant accelerometer arrays:
  Theory, simulation and implementation,'' \emph{Mechatronics}, vol.~23, no.~5,
  pp. 518--529, 2013.

\bibitem{Park2011}
S.~Park and S.~K. Hong, ``Angular rate estimation using a distributed set of
  accelerometers,'' \emph{Sensors}, vol.~11, no.~11, pp. 10\,444--10\,457,
  2011.

\bibitem{Schopp2010}
P.~Schopp, L.~Klingbeil, C.~Peters, and Y.~Manoli, ``Design, geometry
  evaluation, and calibration of a gyroscope-free inertial measurement unit,''
  \emph{Sensors and Actuators A: Physical}, vol. 162, no.~2, pp. 379 -- 387,
  Aug. 2010.

\bibitem{Parsa2004}
K.~Parsa, J.~Angeles, and A.~K. Misra, ``Rigid-body pose and twist estimation
  using an accelerometer array,'' \emph{Archive of Applied Mechanics}, vol.~74,
  no. 3-4, pp. 223--236, 2004.

\bibitem{Cardou2008}
P.~Cardou and J.~Angeles, ``Angular velocity estimation from the angular
  acceleration matrix,'' \emph{Journal of Applied Mechanics}, vol.~75, 2008.

\bibitem{Qin2009}
Z.~Qin, L.~Baron, and L.~Birglen, ``Robust design of inertial measurement units
  based on accelerometers,'' \emph{J. Dyn. Sys., Meas., Control}, vol. 131,
  no.~3, p. 031010, 2009.

\bibitem{Parsa2007}
K.~Parsa, T.~A. Lasky, and B.~Ravani, ``Design and implementation of a
  mechatronic, all-accelerometer inertial measurement unit,'' \emph{{IEEE/ASME}
  Trans. Mechatronics}, vol.~12, no.~6, pp. 640--650, Dec. 2007.

\bibitem{Liu2014}
C.~Liu, S.~Zhang, S.~Yu, X.~Yuan, and S.~Liu, ``Design and analysis of
  gyro-free inertial measurement units with different configurations,''
  \emph{Sensors and Actuators A: Physical}, vol. 214, pp. 175--186, Aug. 2014.

\bibitem{Edwan2011}
E.~Edwan, S.~Knedlik, and O.~Loffeld, ``Constrained angular motion estimation
  in a gyro-free {IMU},'' \emph{{IEEE} Trans. Aerosp. Electron. Syst.},
  vol.~47, no.~1, pp. 596--610, January 2011.

\bibitem{Jiang2012}
C.~Jiang, L.~Xue, H.~Chang, G.~Yuan, and W.~Yuan, ``Signal processing of {MEMS}
  gyroscope arrays to improve accuracy using a 1st order {M}arkov for rate
  signal modeling,'' \emph{Sensors}, vol.~12, no.~2, pp. 1720--1737, 2012.

\bibitem{Yuksel2011}
Y.~Yuksel and N.~El-Sheimy, ``An optimal sensor fusion method for skew
  redundant inertial measurement units,'' \emph{Journal of Applied Geodesy},
  vol.~5, pp. 99--115, 2011.

\bibitem{Xue2012}
L.~Xue, C.-Y. Jiang, H.-L. Chang, Y.~Yang, W.~Qin, and W.-Z. Yuan, ``A novel
  {K}alman filter for combining outputs of {MEMS} gyroscope array,''
  \emph{Measurement}, vol.~45, no.~4, pp. 745 -- 754, 2012.

\bibitem{Bancroft2011}
J.~Bancroft and G.~Lachapelle, ``Data fusion algorithms for multiple inertial
  measurement units,'' \emph{Sensors}, vol.~12, pp. 3720--3738, 2011.

\bibitem{Jafari2014}
M.~Jafari, T.~Najafabadi, B.~Moshiri, S.~Tabatabaei, and M.~Sahebjameyan,
  ``{PEM} stochastic modeling for {MEMS} inertial sensors in conventional and
  redundant {IMU}s,'' \emph{{IEEE} Sensors J.}, vol.~14, no.~6, pp. 2019--2027,
  June 2014.

\bibitem{Guerrier2009}
S.~Guerrier, ``Improving accuracy with multiple sensors: Study of redundant
  {MEMS-IMU/GPS} configurations,'' in \emph{Proc. of the 22nd International
  Technical Meeting of The Satellite Division of the Institute of Navigation
  (ION GNSS)}, Savannah, GA, Sep. 2009.

\bibitem{Bancroft2009}
J.~Bancroft, ``Multiple {IMU} integration for vehicular navigation,'' in
  \emph{Proc. of the 22nd International Technical Meeting of The Satellite
  Division of the Institute of Navigation (ION GNSS)}, Savannah, GA, Sep. 2009.

\bibitem{NSH2014}
J.~O. Nilsson, I.~Skog, and P.~H\"{a}ndel, ``Aligning the forces -- eliminating
  the misalignments in {IMU} arrays,'' \emph{{IEEE} Trans. Instrum. Meas.},
  vol.~63, no.~10, pp. 2498--2500, Oct. 2014.

\bibitem{Jekeli2001}
C.~Jekeli, \emph{Inertial Navigation Systems with Geodetic Applications}.\hskip
  1em plus 0.5em minus 0.4em\relax Walter de Gruyter, 2001.

\bibitem{Stoica1995}
P.~Stoica and A.~Nehorai, ``\BIBforeignlanguage{English}{On the concentrated
  stochastic likelihood function in array signal processing},''
  \emph{\BIBforeignlanguage{English}{Circuits, Systems and Signal Processing}},
  vol.~14, no.~5, pp. 669--674, 1995.

\bibitem{Kay}
S.~Kay, \emph{Fundamentals of Statistical Signal Processing: Estimation
  Theory}.\hskip 1em plus 0.5em minus 0.4em\relax Prentice Hall, 1993.

\bibitem{El-Diasty2007}
M.~El-Diasty, A.~El-Rabbany, and S.~Pagiatakis, ``Temperature variation effects
  on stochastic characteristics for low-cost {MEMS}-based inertial sensor
  error,'' \emph{Meas. Sci. Technol.}, vol.~18, no.~11, pp. 3321--3328, Sep.
  2007.

\bibitem{Tao2011}
T.~Li and A.~Nehorai, ``Maximum likelihood direction finding in spatially
  colored noise fields using sparse sensor arrays,'' \emph{{IEEE} Trans. Signal
  Process.}, vol.~59, no.~3, pp. 1048--1062, Mar. 2011.

\bibitem{Lapinski2009}
M.~Lapinski, E.~Berkson, T.~Gill, M.~Reinold, and J.~Paradiso, ``A distributed
  wearable, wireless sensor system for evaluating professional baseball
  pitchers and batters,'' in \emph{Proc. International Symposium on Wearable
  Computers (ISWC)}, Linz, Austria, Sep. 2009, pp. 131--138.

\bibitem{Camarillo2013}
D.~Camarillo, P.~Shull, J.~Mattson, R.~Shultz, and D.~Garza, ``An instrumented
  mouthguard for measuring linear and angular head impact kinematics in
  american football,'' \emph{Annals of Biomedical Engineering}, vol.~41, no.~9,
  pp. 1939--1949, 2013.

\bibitem{Gorman1990}
J.~Gorman and A.~Hero, ``Lower bounds for parametric estimation with
  constraints,'' \emph{{IEEE} Trans. Inf. Theory}, vol.~36, no.~6, pp.
  1285--1301, Nov. 1990.

\bibitem{SNH2014b}
I.~Skog, J.-O. Nilsson, and P.~H\"{a}ndel, ``Pedestrian tracking using an {IMU
  }array,'' in \emph{Proc. IEEE International Conference on Electronics,
  Computing and Communication Technologies (CONECCT)}, Bangalore, India, Jan.
  2014.

\end{thebibliography}

\begin{IEEEbiography}[{\includegraphics[width=1in,height=1.25in,clip,keepaspectratio]{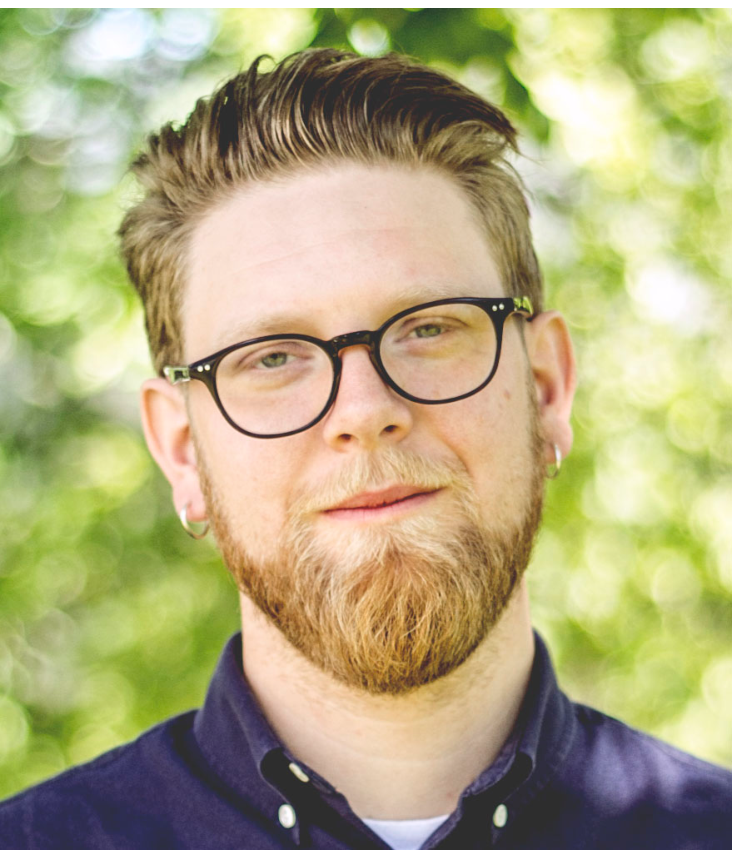}}]
{Isaac Skog}(S'09-M'10) received the BSc and MSc degrees in Electrical Engineering from the KTH Royal Institute of Technology, Stockholm, Sweden, in 2003 and 2005, respectively. In 2010, he received the Ph.D. degree in Signal Processing with a thesis on low-cost navigation systems. In 2009, he spent 5 months at the Mobile Multi-Sensor System
research team, University of Calgary, Canada, as a visiting scholar and in 2011 he spent 4 months at the Indian Institute of Science (IISc), Bangalore, India, as a visiting scholar. He is currently a Researcher at KTH coordinating the KTH Insurance Telematics Lab. He was a recipient of a Best Survey Paper Award by the IEEE Intelligent Transportation Systems Society in 2013.
\end{IEEEbiography}

\begin{IEEEbiography}[{\includegraphics[width=1in,height=1.25in,clip,keepaspectratio]{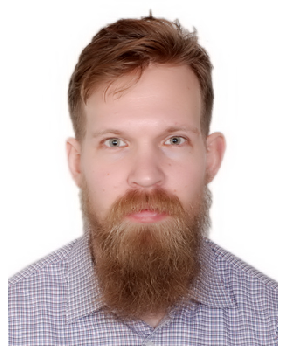}}]
{John-Olof Nilsson} (M'14) received the M.Sc. degree in Engineering Physics from Royal Institute of Technology, Stockholm, Sweden, in 2008. In 2013, he received the Ph.D. degree in Signal Processing with a thesis on infrastructure-free pedestrian localization. In 2011, he spent 4 months at the Indian Institute of Science (IISc), Bangalore, India as a visiting scholar and in 2014 he spent 3 months at the Indian Institute of Technology (IIT) Kanpur, India, as a visiting Scholar. He was a recipient of the Best Demonstration Award at the IEEE 2014 International Conference on Indoor Positioning and Indoor Navigation, Busan, Korea.
\end{IEEEbiography}

\begin{IEEEbiography}[{\includegraphics[width=1in,height=1.25in,clip,keepaspectratio]{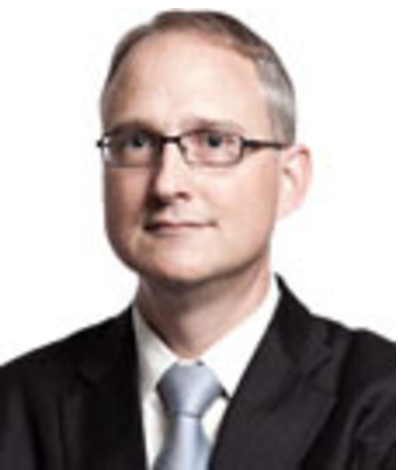}}]
{Peter H\"{a}ndel}(S'88-M'94-SM'98) received a Ph.D. degree from Uppsala University, Uppsala, Sweden, in 1993. From 1987 to 1993, he was with Uppsala University. From 1993 to 1997, he was with Ericsson AB, Kista, Sweden. From 1996 to 1997, he was a Visiting Scholar with the Tampere University of Technology, Tampere, Finland. Since 1997, he has been with the KTH Royal Institute of Technology, Stockholm, Sweden, where he is currently a Professor of Signal Processing and the Head of the Department of Signal Processing. From 2000 to 2006, he held an adjunct position at the Swedish Defence Research Agency. He has been a Guest Professor at the Indian Institute of Science (IISc), Bangalore, India, and at the University of G\"avle, Sweden. He is a co-founder of Movelo AB. Dr. H\"andel has served as an associate editor for the IEEE TRANSACTIONS ON SIGNAL PROCESSING. He was a recipient of a Best Survey Paper Award by the IEEE Intelligent Transportation Systems Society in 2013.
\end{IEEEbiography}

\begin{IEEEbiography}[{\includegraphics[width=1in,height=1.25in,clip,keepaspectratio]{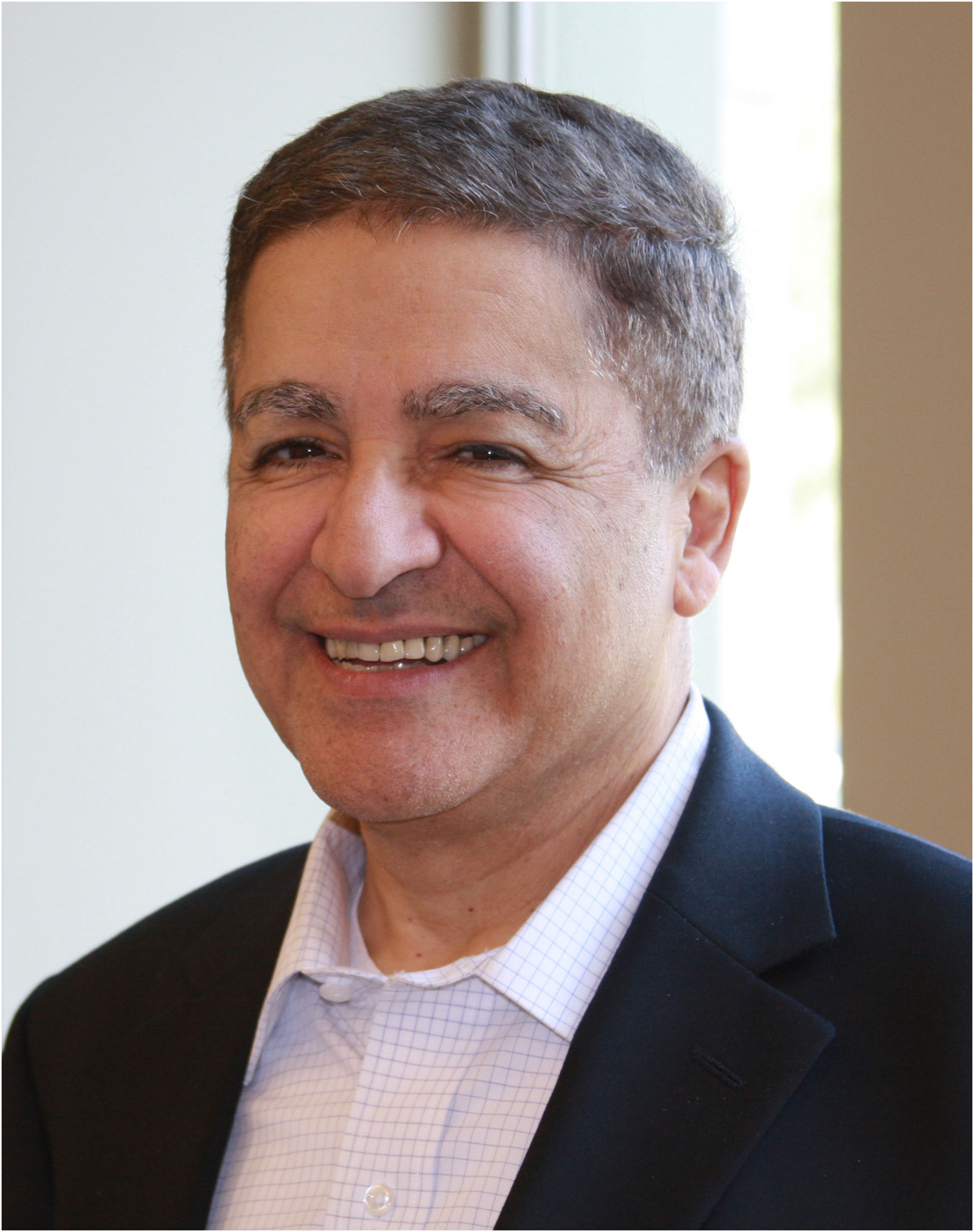}}]
{Arye Nehorai} (S'80--M'83--SM'90--F'94) is the Eugene and Martha Lohman Professor and Chair of the Preston M.\ Green Department of Electrical and Systems Engineering (ESE), Professor in the Department of Biomedical Engineering (by courtesy) and in the Division of Biology and Biomedical Studies (DBBS) at Washington University in St.\ Louis (WUSTL). He serves as the Director of the Center for Sensor Signal and Information Processing at WUSTL.  Under his leadership as department chair, the undergraduate enrollment has more than tripled in the last four years. Earlier, he was a faculty member at Yale University and the University of Illinois in Chicago. He received both B.Sc.\ and M.Sc.\ degrees from the Technion, Israel and a Ph.D.\ from Stanford University, California.

Dr.\ Nehorai served as Editor-in-Chief of the {\em IEEE Transactions on Signal Processing}\/ from 2000 to 2002. From 2003 to 2005 he was the Vice President (Publications) of the IEEE Signal Processing Society (SPS), the Chair of the Publications Board, and a member of the Executive Committee of this Society. He was the founding editor of the special columns on Leadership Reflections in the {\em IEEE Signal Processing Magazine}\/ from 2003 to 2006.

Dr.\ Nehorai received the 2006 IEEE SPS Technical Achievement Award and the 2010 IEEE SPS Meritorious Service Award. He was elected Distinguished Lecturer of the IEEE SPS for a term lasting from 2004 to 2005. He received several best paper awards in IEEE journals and conferences. In 2001 he was named University Scholar of the University of Illinois. Dr.\ Nehorai was the Principal Investigator of the Multidisciplinary University Research Initiative (MURI) project entitled Adaptive Waveform Diversity for Full Spectral Dominance from 2005 to 2010. He has been a Fellow of the IEEE since 1994, a Fellow of the Royal Statistical Society since 1996, and a Fellow of AAAS since 2012.
\end{IEEEbiography}

\end{document}